\documentclass[a4paper,12pt]{article}
\usepackage{epsfig}
\usepackage{graphicx}
\usepackage{subfigure}
\begin{document}

\title{\textbf{Q-Ball Scattering on Barriers and Holes in 1 and 2 Spatial Dimensions}}
\author{Jassem H. Al-Alawi\thanks{e-mail address:J.H.Al-Alawi@durham.ac.uk} and Wojtek J. Zakrzewski\thanks{email address: W.J.Zakrzewski@durham.ac.uk}
\\ Department of Mathematical Sciences,University of Durham, \\
 Durham DH1 3LE, UK\\
}

\date{\today}

\maketitle

\begin{abstract}
We discuss various scattering properties of non-topological solitons, Q-balls, on potential obstructions in $\left(1+1\right)$ and $\left(2+1\right)$ dimensions. These obstructions, barriers and holes, are inserted into the potential of the theory via the coupling parameter, ${\it ie}$ $\tilde\lambda$, that is effective only in a certain region of space. When $\tilde\lambda>1$ the obstruction is a barrier and when $0<\tilde\lambda<1$ the obstruction is a hole. The dynamics of Q-balls on such obstructions in $\left(1+1\right)$ dimensions is
shown to be very similar to that of topological solitons provided that the
Q-balls are stable.

 In $\left(2+1\right)$ dimensions, numerical simulations have shown some differences from the dynamics
of topological solitons. We discuss these differences in some detail.
%The Q-ball behaviour in two dimensions when only the tail of the Q-ball is involved in the interaction with obstructions is similar to some known scatterings in physics,{\it i.e} Rutherford scattering in the case of barrier and electromagnetic scattering of a point charge from a source containing collective of point charges in the case of holes. The similarity is due to having the same parameters that are important in the scattering behaviour of Q-balls and these scatterings.
\end{abstract}

\section{\textbf{Introduction}}

 Motivated by our earlier work on the scattering of topological solitons on potential  obstructions in a class of models [1,2] we have decided to look at the generality of these results. To do this we have decided 
to investigate the scattering properties of non-topological solitons so in this paper we look
at the scattering of Q-balls on various potential obstructions.

 First we recall that Q-balls are non-topological soliton fields, first introduced by S. Coleman in [3]. They arise in some classes of self-interacting complex scalar fields. Unlike topological solitons, whose stability is ensured by the presence of a conserved topological charge, Q-balls carry a conserved Noether charge associated with the global $\,U\left(1\right)$ symmetry.  This conserved but non-topological charge stabilises the Q-balls. Also, Q-balls have a time-dependent rotating phase and the conserved Noether charge is related to the velocity
of the change of this phase. These features result in a much more complicated  scattering properties of these solitons, leading to such phenomena as, for example, the charge fission during the scattering process.

 In this paper we investigate the Q-ball dynamics in $\left(1+1\right)$ and $\left(2+1\right)$ dimensions, mainly 
when the Q-balls are scattered on various obstructions in the form of potential barriers or holes. First 
we discuss the conditions of the stability of the Q-balls. Then\index{\footnote{}} we report
the results of our numerical simulations which have revealed that Q-balls behave like topological solitons only 
for the values of the parameters of the theory that guarantee this stability throughout the scattering.
 When this condition is satisfied the dynamics of the scattering on barriers and holes in $\left(1+1\right) $ dimensions is similar to what was seen for topological solitons [1,2], {\it ie}  very little radiation is given off during the scattering. The scattering on barriers results in transmission 
or reflection, {\it ie} very much like that of a small particles. In the scattering on holes, like for
topological solitons, we can have transmissions, trappings and reflections. As discussed before [1-2, 4-5] the
trappings correspond to the case of solitons which, while in the hole, lose energy through radiation and the reflections correspond to the very coherent interaction of solitons with this radiation.
We find similar properties of the scattering of Q-balls except that this time they also generate
further small Q-balls. This scattering is discussed in detail in sections 3 and 4.

% However, the questions regarding the behaviour on holes even in the case of Q-balls are still not completely understood within a theory that is fully classical and relativistic. A Q-ball as is the topological soliton will have a critical velocity above which the Q-ball will be able to get out of the hole and below which it gets trapped inside the hole losing very tiny amount of energy in form of radiation. The behaviour is far challenging when the Q-ball gets reflected from a hole in a behaviour that is non-classical. This is the behaviour that remains challenging to a theory which is classical in topological soliton case. In section 3 and section 4 we have studied in details the scattering on barriers and holes. 

In section 5 we discuss the scattering of solitons in $\left(2+1\right)$ dimensions. We find that, again, the scattering often resembles the scattering of small particles but the properties of this scattering
depend crucially on the parameters of the theory. However, there are also small differences which we discuss 
in detail. The paper ends with a short section presenting our conclusions.

% In this paper a further interesting dynamics has been manifested when the scattering is studied in $\left(2+1\right)$ dimensions. The scattering in two dimensions has given a further confirmation of the particle nature of Q-balls only when the tail of the Q-balls are the portions that interact with obstructions. The numerical simulations have shown that the scattering of Q-balls in two dimensions resembles some known scattering in physics,{\it i.e.} Rutherford and electromagnetic scatterings as we will see in sections 5.1 and 5.2. The scattering of Q-ball on barriers resembles Rutherford scattering and the scattering on holes resembles electromagnetic scattering. By resemblance we mean that the parameters that play important role in the Q-ball scattering are the same ones that are important in the Rutherford and electromagnetic scatterings. 

\section{\textbf{Q-balls in $D$ Spatial Dimensions}}

We consider a complex scalar field $\Phi$ in $D$ spatial dimensions [6] whose Lagrangian density is given by
\begin{equation}
\mathcal{L}=\frac{1}{2}\partial_{\mu}\Phi\partial^{\mu}\Phi^{\ast}-\,U\left(\vert\Phi\vert\right),
\label{lag}
\end{equation}
where the indices $\mu$ and $\nu$ run from 0 to $D$. The energy momentum tensor takes the form:
	
\begin{equation}
\,T_{\mu\nu}=\partial_{\mu}\Phi\partial_{\nu}\Phi^{\ast} - \,g_{\mu\nu}\mathcal{L},
\end{equation}
	where $\,g_{\mu\nu}$ is the space-time metric with signature $\left(+,--\dots\right)$. This Lagrangian has been so chosen that it is invariant under the global $\,U\left(1\right)$ transformation and the conserved Noether current,$\,j^{\mu}$, associated with this symmetry, is given by:

\begin{equation}
\,j^{\mu}=\frac{1}{2{\it i}}\left(\Phi^{\ast}\partial^{\mu}\Phi-\Phi\partial^{\mu}\Phi^{\ast}\right),
\end{equation}

The corresponding conserved Noether charge,$\,Q,$ is then

\begin{equation}
\,Q=\int\,j^{0}\,d^{D}\,x.
\end{equation}

The total energy and momentum are given, respectively, by
\begin{equation}
\,E=\int\,T_{00}\,d^{D}\,x,
\end{equation}

\begin{equation}
\,P_i=\int\,T_{0\it i}\,d^{D}\,x,
\end{equation}
where $\it i$ is the spatial index which runs from 1 to $D$.

The stationary Q-ball located at the origin is described by:
 \begin{equation}
 \label{field}
\Phi=\,e^{\,i\omega\,t}\,f\left(\,r\right),
\end{equation}
where $\omega$ is the internal rotation frequency and $\,f\left(\,r\right)$ is the real radial profile function 
which satisfies the equation of motion which follows from (1), namely:
\begin{equation}
\label{eone}
\frac{\,d^{2}f}{\,dr^{2}}=\frac{\left(1-D\right)}{r}\frac{\,df}{\,dr}-\omega^{2}\,f+\,U^{\prime}\left(\,f\right),
\end{equation}
where $D$ is the number of spatial dimensions.

For the consistency of the equations and in order that the Q-ball has a finite energy,
the profile function $f$, has to satisfy the following
 boundary conditions
 \begin{equation}
\frac{\,df\left(0\right)}{\,dr}=0
\end{equation}
and 
\begin{equation}
\,f \rightarrow 0 \,\quad  \hbox{as}\quad  \,\vert\,r\vert \rightarrow \infty.
\end{equation}

The corresponding field (7) is then called the stationary Q-ball.

As the Lagrangian (1) is Lorentz invariant  the field (7) can be boosted with any velocity up to the speed of light.

 Note that the equation for $f$, (8), can be treated as a mechanical analogue  describing a point particle moving in an effective potential with friction if $\,f$ is interpreted as the position of the particle and $r$ as the time coordinate [6].
  The effective potential in which this particle moves is then
\begin{equation}
\label{etwo}
 \,U_{eff}\left(f\right)=\frac{1}{2}\omega^{2}\,f^{2}-\,U\left(f\right).
\end{equation}

 The `friction' term $\frac{1-\,D}{r}$ becomes important for high $D$ or small $r$.

\subsection{\textbf{Q-balls in $\left(1+1\right)$ Dimensions}}

First let us choose our potential $U(f)$.  We want it to be such that it gives rise to an exact Q-ball solution in $\left(1+1\right)$ dimensions. This potential should have an absolute minimum when $\,U\left(0\right)=0 $. Then  $\Phi=0 $ would be the ground state of the theory. Moreover, we require that the symmetry of the theory remains unbroken.

However, looking at (11) we note that there are further conditions on the parameters
of the Q-ball ({\it ie} $\omega$).  To see this we observe that the Q-ball field exists  provided that $\,f=0$ corresponds to a local maximum of the effective potential {\it ie} $\,U^{\prime\prime}_{eff}\left(0\right)=\left[2\,U/\,f^{2}\right]_{f=0}<0 $. This leads to an upper bound on the frequency $\omega=\omega_{+}$. Also,  the local minimum of $ \left[2\,U\left(\,f\right)/\,f^{2}\right]$ must be attained at some positive value of $\,f=\,f_{0}$ and this requirement puts a lower bound on $\omega$  {\it ie} $\omega=\omega_{-}$.
\begin{center}
 \begin{equation}
\index{}\left[2\,U\left(\,f_{0}\right)/\,f_{0}^{2}\right]  <\omega < \,U^{\prime\prime}\left(0\right).
 \end{equation}
\end{center}

 Thus the Q-ball field exists for $\omega$ in this range, {\it ie} $ \omega_{-}<\omega<\omega_{+}$. Of course, as (11) is quadratic 
 in $\omega$, we can also take negative values of $\omega$ -  the corresponding fields are called anti-Q-balls.  However, the existence of a Q-ball does not guarantee its stability. For a Q-ball field configuration to be stable we have to require that 
  $\,E\left(\,Q\right)< \,m\,Q$, where $m$ is the mass of the scalar field. 

The potential we have chosen to use has the form [7]:

\begin{equation}
\label{ethree}
 \,U\left(\vert\Phi\vert\right)=2\vert\Phi\vert^{2} -2 \vert\Phi\vert^{4} +\vert\Phi\vert^{6}.
 \end{equation}
 
With this potential  Q-ball solutions exist for $\sqrt{2}<\omega<2$. Moreover, these
Q-balls are stable as $m=2$.

Next we insert our obstructions. Following our earlier work [1-2] on the soliton scattering
 on obstructions we introduce a potential parameter that perturbs the Q-ball 
  only in a certain region of space. The Q-ball, as it moves towards this region, will 
  only experience this perturbation when its profile function $f$
  in non-zero in this region.  
  We introduce the perturbation in the form of  potential barriers and holes, and hence we modify the potential $U(\Phi)$ to become
 
%where the potential $\,\tilde V\left(\vert\varphi\vert\right)$ is taken in the form
\begin{equation}
\,\tilde U\left(\vert\Phi\vert\right)=\tilde \lambda\,U\left(\vert\Phi\vert\right),
\end{equation}
in which  $\tilde \lambda=1+\lambda\left(\,x\right)$.
 
Thus $\lambda\left(\,x\right)$ is the extra potential parameter which has been inserted into the potential $\,\tilde U\left(\vert\Phi\vert\right)$ to take into account the effects of  obstructions, holes and barriers, and so is non-zero only in a certain region of space.

In our case, we have put the obstruction around the origin ({\it ie} $x=0$) and
so we have taken
\begin{center}
$\lambda\left(\,x\right)=\cases{ 0 & $\vert\,x\vert> 10 $ \cr
                                 \lambda_{0}=\hbox{constant} & $\vert\,x\vert\leq 10. $ \cr} $

\end{center}

 The Q-ball in our model is really quite sizeable ({\it ie} the solution of (8) for the 
 potential (13) is nonzero for a sizeable region of $x$)
  so the obstruction has been made quite large to make sure that the Q-ball size fits into
  it quite easily; {\it ie} is significantly larger than the size of the Q-ball.
  
The equation of motion, in the presence of the obstruction, now takes the form
\begin{equation}
 \ddot\Phi-\Phi^{\prime\prime}+2\Phi\tilde\lambda\left(2-4\vert\Phi\vert^{2}+3\vert\Phi\vert^{4}\right)=0
\end{equation}

Using the Q-ball ansatz, the field equation for the profile function, $\,f\left(\,x\right)$ 
now takes the form:
\begin{equation}
\label{efive}
 \,f^{\prime\prime}+\left(\omega^{2}-4\tilde\lambda\right)+ 8\tilde\lambda\,f^{3}-6\tilde\lambda\,f^{5}=0. 
\end{equation}

This profile equation (16) can only  be solved exactly when $\lambda\left(\,x\right)=\lambda_{0}$=constant.  Then with the help of [8]  we note that $f$ is given by

\begin{equation}
\,f\left(\,x\right)=\left(\frac{1}{\tilde\lambda}\right)^{\frac{1}{4}}\left[\frac{4\tilde\lambda-\omega^{2}}{2\sqrt{\tilde\lambda}+\sqrt{2\omega^{2}-4\tilde\lambda}\,\cosh\left(2\,x\sqrt{4\tilde\lambda-\omega^{2}}\right)}\right]^{\frac{1}{2}}.
\end{equation}

As the theory is Lorentz invariant this solution can be be boosted or shifted by a spatial translation. Then, the Q-ball solution becomes 
\begin{equation}
\label{boosted}
\,f\left(\,x\right)=\left(\frac{1}{\tilde\lambda}\right)^{\frac{1}{4}}\left[\frac{4\tilde\lambda-\omega^{2}}{2\sqrt{\tilde\lambda}+\sqrt{2\omega^{2}-4\tilde\lambda}\,\cosh\left(2\gamma\sqrt{4\tilde\lambda-\omega^{2}}\left(\,x-\,x_{0}-\,u\,t\right)\right)}\right]^{\frac{1}{2}}
\end{equation}
where $\,x_{0}$ is the position of the  Q-ball, $\,u$ its velocity and $\gamma$ is the usual relativistic factor.

However, $\tilde\lambda\ne$ constant and so this expression does not solve (16) for all values of the parameters. On the other hand our $f(x)$, (18), is strongly localised and so if the Q-ball is far away from 
the obstruction the solution to (16) is approximately given by (18) 
and so (18) can be used for the initial condition of our numerical
simulation of (15).

Note that when the Q-ball is far enough from the obstruction $\tilde\lambda=1.0$ and 
(18) becomes

\begin{equation}
\,f\left(\,x\right)=\left[\frac{4-\omega^{2}}{2+\sqrt{2\omega^{2}-4}\,\cosh\left(2\gamma\sqrt{4-\omega^{2}}\left(\,x-\,x_{0}-\,u\,t\right)\right)}\right]^{\frac{1}{2}}
\end{equation}

The associated Q-ball Noether charge and the energy become

\begin{equation}
 \,Q\,=\, \omega\gamma\,I_{2}
\end{equation}

 \begin{equation}
\,E\,=\,\frac{1}{2}\gamma^{2}\left(\,u^{2}+1\right)\,I_{\,x}+\frac{1}{2}\omega^{2}\gamma^{2}\left(\,u^{2}+1\right)\,I_{2}+2\,I_{2}-2\,I_{4}+\,I_{6},
 \end{equation}
where 

$\,I_{2}=\int_{-\infty}^{\infty}\,f^{2}\,dx=\sqrt{2}\omega\tanh^{-1}\left[{\frac{2-\omega^{\prime}}{2+\omega^{\prime}}}\right]^\frac{1}{2}$
and $\omega^{\prime}=\sqrt{2\omega^{2}-4}$.

Moreover,
\begin{center}
$\,I_{4}=\int_{-\infty}^{\infty}\,f^{4}\,dx= -\frac{4-\omega^{2}}{2}+\,I_{2}$,

$\,I_{6}=\int_{-\infty}^{\infty}\,f^{6}\,dx= \frac{-3\sqrt{4-\omega^{2}}}{4}+\frac{\left(\omega^{2}+2\right)}{4}\,I_{2}$,\\
$\,I_{\,x}=\int_{-\infty}^{\infty}\,f^{\prime^{2}}\,dx=\frac{4-\omega^{2}}{2}-\frac{\omega^{\prime^{2}}}{4}\,I_{2}.$
\end{center}

We see that both $Q$ and $E$ depend on $\omega$ and so, in follows, we call them 
$Q_{\omega}$ and $E_{\omega}$.

Thus the Q-ball Noether charge is 

\begin{equation}
\,Q_{\omega}=\sqrt{2}\omega\gamma\tanh^{-1}\left[{\frac{2-\omega^{\prime}}{2+\omega^{\prime}}}\right]^\frac{1}{2}
\end{equation}
and the energy becomes

\begin{equation}
\,E_{\omega}=\frac{1}{4}\left(\gamma^{2}\left(\,u^{2}+1\right)+1\right)\left[\sqrt{4-\omega^{2}}+\left(\omega^{2}+2\right)\frac{\,Q_{\omega}}{\omega}\right].
\end{equation}

When $u\,=\,0$ ({\it ie} the Q-ball is at rest) the energy becomes its mass and  is given by 

\begin{equation}
 \,M_{\omega}=\frac{\sqrt{4-\omega^{2}}}{2}+\frac{\left(\omega^{2}+2\right)}{2}\frac{\,Q_{\omega}}{\omega}.
\end{equation}

Looking at these formulae we observe that as $\omega \approx \omega_{+}$,  $\,Q_{\omega}$ and $\,E_{\omega} \rightarrow 0$ and when $\omega\approx\omega_{-}$ then $\,Q_{\omega} $ and $\,E_{\omega}\rightarrow \infty$. Fig. 1 shows that $E_{\omega} $ and $Q_{\omega}$ are monotonically decreasing functions of $\omega$. Furthermore, as $\omega$ increases the charge decreases and in the limit as $\omega\approx\omega_{+}$ the charge, $\,Q_{\omega}\rightarrow 0$. In this upper bound limit the energy-charge ratio approaches the mass of the scalar field, {\it ie} $\,m=2$. In the lower bound limit, {\it ie} when $ \omega\approx\omega_{-}$, the ratio $\frac{\,E_{\omega}}{\,Q_{\omega}}\rightarrow \omega_{-} $. Fig. 2 shows clearly that the energy-charge ratio is below the mass term of the scalar field when the velocity of the Q-ball is small [9]. When $\,u=0.0$ the energy-charge ratio is always below the stability line and only intersects  this line when $\omega\sim\omega_{+}$. Thus, the Q-balls
are stable for small velocities and so are prevented from decaying into a number of smaller Q-balls. However, when the Q-ball acquires higher speeds the absolute stability condition can be violated; Q-balls can still be long-lived and  will, eventually, decay into a number of smaller Q-balls [7].  Fig. 2  shows that as $\omega$ increases and so  $\,E_{\omega}$ and $\,Q_{\omega}$  decrease the stability condition becomes more violated for higher velocities. 

When an obstruction is present in a certain region of space the mass term gets modified by the perturbation factor $\tilde\lambda$ {\it ie}  $\,m=2\sqrt{\tilde\lambda}$.
Thus, the stability line  shown in fig. 2 gets shifted and the amount of the shift depends on the magnitude of the perturbation. In the case of a barrier the stability line gets shifted upwards and downwards in the case of a hole. Consequently, the Q-ball is less stable in the presence of holes. This is what we have observed in our simulations.

 This scattering on holes is very delicate since if we want the Q-ball to be stable against fission, as the Q-ball reaches the hole, the hole must be relatively shallow.

 So, when the potential is lowered by a positive factor (but less than one), {\it ie} $0 <\tilde\lambda< 1$, the  tail of the Q-ball may develop some small Q-balls in the hole even when the Q-ball has a vanishing velocity. On the other hand, the Q-ball stability is not affected by the height of the barrier in cases when the velocity is small. In fact, within some limits, the higher the barrier the more stable is the Q-ball scattering as the stability line is shifted further upwards. However, we have to be careful; the Q-ball may also become unstable if
the barrier is too high. % And so, when the barrier is high within the stability limit that is rescaling the potential by a positive coefficient greater than one,{\it i.e} $\tilde\lambda>1$ the Q-ball will be more stable during the scattering even when the Q-ball is sent at a high speed. 

\begin{figure}
\begin{center}
\includegraphics[angle=270, width=10cm]{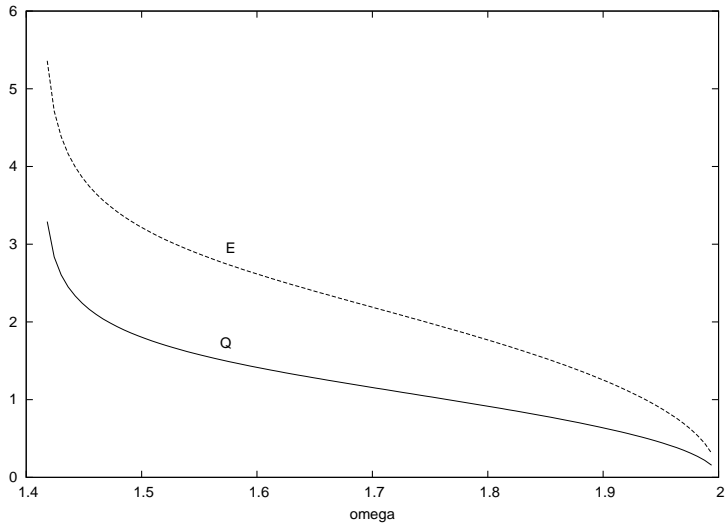}
\caption{The energy $\,E_{\omega}$ and the charge $\,Q_{\omega}$ as a function of   $\omega$. }
\end{center}
\end{figure}

\begin{figure}
\begin{center}
\includegraphics[angle=270, width=12cm]{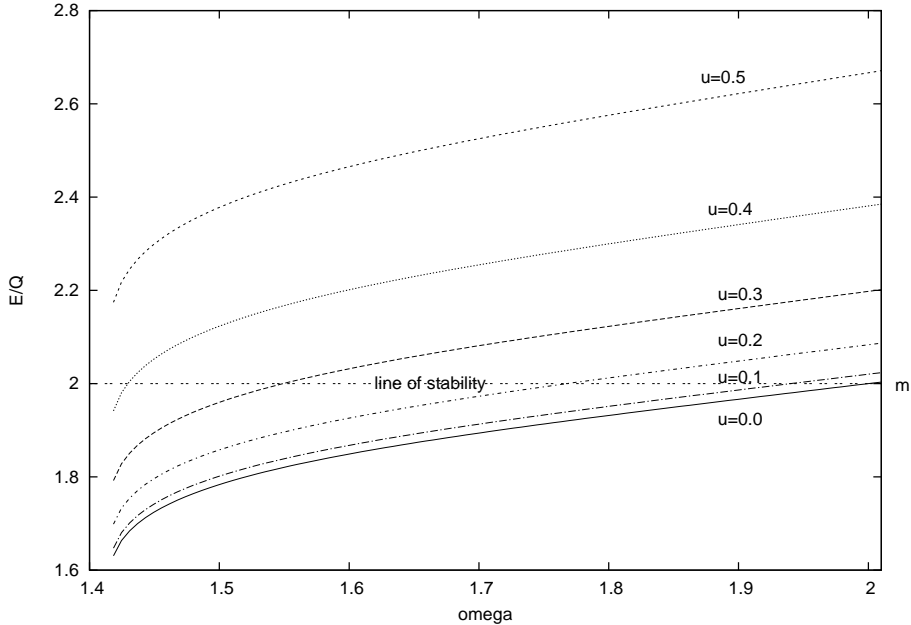}
\caption{The Energy-charge ratio $E$/ $Q$  as a function of $\omega$ for various velocities.}
\end{center}
\end{figure}

In next two sections we discuss the dynamics of Q-balls as they meet an obstruction located around $\,x=0$. In order to understand it we had to perform some numerical simulations.
These simulations were performed using the 4th order Runge - Kutta method of simulating the time evolution. We used 12001 points with the lattice spacing of  $dx=0.01$. Hence, the lattice extended from -60 to 60 in the $x$-direction. The time step was chosen to be $dt=0.0025$. In our work we used the absorbing boundary conditions. 
%\begin{equation}
% \,f\left(0\right)=\sqrt{\frac{4-\omega^{2}}{2+\sqrt{2\omega^{2}-4}}}
%\end{equation}

%expand $\,f=\sum\,a_{i}\psi_{i}$, where $\psi_{i}$ is orthonormalized eigenvectors. The stability is related to the eigenvalue problem

%\begin{equation}
% \^{H}\psi_{i}=\,c_{i}\psi_{i}
%\end{equation}

%Hence, the absolute stability condition is satisfied and the Q-ball is preserved from decaying into a number of smaller Q-balls.  This condition would be viloated when the Q-ball is moving at a high speed and during the Q-ball scattering on obstruction; and so the Q-ball will fission into a number of smaller Q-balls as we will see later in the following sections. The more energy the Q-ball has the more stable it is against fission. 

% TEST WHAT YOU HAVE WROTE %%%%%%%%%%%%%%%%%%%%%%%%%%%%%%%%%%%
% insert a Table showing in that table u,w,E/Q FOR ALL W AND VARIOUS u%%%%%%%%%%%%%%%%%%%%%%%%%%%%%%555555555

%Q-ball anstaz has the form

%We have chosen a typical simple potential in field theories that give rise to Q-ball solutions,

%\begin{equation}
% \,U=\,f^{2}\left(1+\left(1-\,f^{2}\right)^{2}\right).
%\end{equation}

%The stationary exact solution is given by

%\begin{equation}
% \,f\left(\,x,\,t\right)=\sqrt{\frac{4-\omega^{2}}{2+\sqrt{2\omega^{2}-4}\,cosh\left(2\gamma\sqrt{4-\omega^{2}}\left(\,x-\,x_{0}-\,u\,t\right)\right)}}
%\end{equation}

%The energy density
%\begin{equation}
% \epsilon=\frac{1}{2}\vert\dot\varphi\vert^{2}+\frac{1}{2}\vert\varphi^{\prime}\vert^{2}+2\vert\varphi\vert^{2}-2\vert\varphi\vert^{4}+\vert\varphi\vert^{6}
%\end{equation}

\section{\textbf{Q-Ball Scattering on Barriers}}

When a Q-ball, or a topological soliton, are far away from an obstruction, to a good approximation, they
can be treated as a point particle following a well defined trajectory. However, when they approach the barrier
the differences in their behaviour begin to arise. These differences stem from the fact that the Q-charge
of the Q-ball is not quantized and that not only Q-balls exist only in a certain range of frequencies
but also there are some conditions on the Q-ball that have to be satisfied for it to be stable.
The work on topological solitons [1-4] has revealed that the speed and the magnitude of the barrier do
 not affect significantly the general dynamics of the solitons. Topological solitons suffer some distortions during their scattering process but they do recover their original shape after the scattering as the scattering is quite elastic. For the Q-balls, however, their dynamics on obstructions (barriers or holes) is connected to their stability. And this stability depends on the velocity, the height of the obstruction and the internal frequency of the Q-balls and thus on their charge.

One major difference between the Q-balls and topological solitons as they scatter on barriers is their
behaviour close to their critical velocities. Critical velocities of the solitons are the lowest velocities above which they can overcome the barrier and  get transmitted  and below which they are reflected. 
All models involving topological solitons that we have studied have shown that the solitons can be treated
as point particles and that such a discussion leads to very accurate estimates
of their critical velocity. There are no restrictions on the speed of the solitons or the height of the barrier. In the case of the Q-balls the situation is different. First of all, the critical velocity cannot be determined without having to impose some restrictions on their speeds and the height of the barrier. This is  due to the possibility of the Q-ball becoming unstable during its interaction with the barrier.

As we have seen from fig. 2 the stability condition is satisfied only for low velocities. When a barrier is present, the stability line is raised and more Q-ball field configurations become stable. 
So to begin with we will consider the scattering on very small barriers.
%To do this we, first of all, have checked that, like for topological solitons, the barriers are repulsive.  

%In order to discuss the Q-ball scattering on barriers we need to have the full spectrum of Q-ball solutions stable. In order to do that we will consider very low barrier heights. 
Before we go on and discuss the scattering dynamics of the Q-ball on such barriers, we have to determine whether the force between the Q-ball and the barrier is repulsive or attractive. To check this we have placed a Q-ball at rest close to the barrier. Fig. 3 presents a trajectory of a Q-ball with $\omega=1.9$ initially placed at rest at $\,x=-15$, {\it ie} five units of distance from a barrier of height 0.01. The 
plot clearly demonstrates that the Q-ball is repelled by the barrier. This repulsion is due to the tail
of the Q-ball which interacts with the barrier. The further away we place Q-ball 
the smaller the tail and hence the weaker the repulsion but the repulsion is always there. So to study the scattering on the barrier we need to give the Q-ball a boost.

%In case of small velocities, Q-balls have shown similar dynamics of barrier scattering to topological solitons. Q-balls move along the line of motion and slow down as they meet the barrier. If they have enough energy to overcome the barrier they will be transmitted otherwise they get reflected. The shape of the Q-ball get squashed at the barrier but they recover their original shape after scattering. There will be a critical velocity above which the Q-ball can be transmitted and below which the Q-ball is reflected. By treating Q-balls as point particles, critical velocities can be calculated nearly exactly,{\it i.e} less than one percent difference, as the numerical values. 

\begin{figure}
\begin{center}
\includegraphics[angle=270, width=10cm]{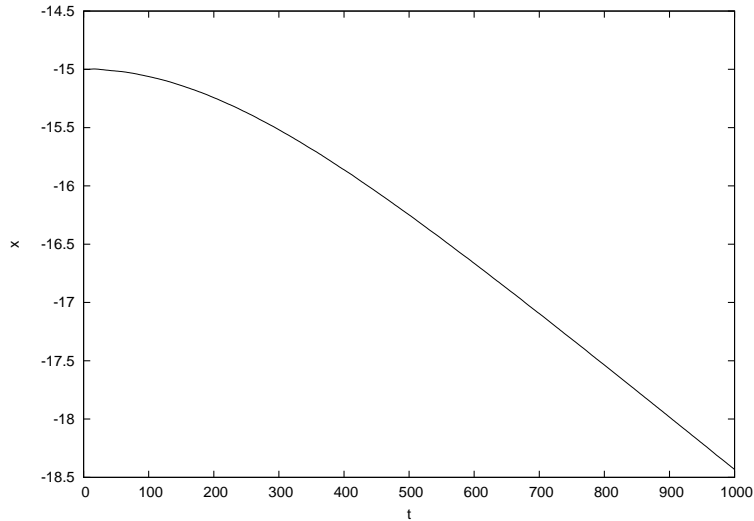}
\caption{The trajectory of a Q-ball solution, $\omega=1.9$, $\,u=0$, placed at $\,x=-15$ from a barrier of 0.01 height}
\end{center}
\end{figure}

We have performed several simulations by boosting the Q-balls with various small velocities.
All these simulations have shown that, like for the topological solitons, the scattering is very elastic.
There is very little radiation sent out and although the shape of the Q-ball changes as it climbs
the barrier - afterwards it returns to its original shape. Looking at the dependence on the frequency ({\it ie} $Q$) of the Q-ball we have noticed that the Q-balls of higher value of $Q$ get closer to the barrier.
However, when we tried to determine the velocity at which they reach the top of the barrier ({\it ie} 
the critical velocity) this dependence on the frequency disappears. Thus for a barrier of height 0.01 
the numerically found value of $u_{cr}$ was $u_{cr}\sim 0.1$.

 As the scattering is very elastic 
we have tried to derive this value in the point particle approximation. In this case we look at the energy of the Q-ball at rest at the top of the barrier and compare it with the 
energy of a moving Q-ball away from the barrier. If the scattering is elastic the equality of 
these energies gives us a lower bound on the critical velocity.

Thus in our case
\begin{center}
$ \,E_{cr}\sim \frac{\,M_{rest}}{\sqrt{1-\,u_{cr}^{2}}}.$
\end{center}
This energy should be very close to the rest mass energy of a Q-ball at the top of a barrier given by
\begin{center}
$ \,E_{cr}\sim \,M_{B}.$
\end{center}
 Thus the critical velocity is given by
\begin{equation}
\,u_{cr}=\sqrt{1-\left(\frac{\,M_{rest}}{\,E_{cr}}\right)^{2}} .
\end{equation}
So, let us use the above formula to estimate the critical velocity. The rest mass energy of the Q-ball with $\omega=1.9$, far away from the potential barrier, is $ \,M_{rest}= 1.2528$ while its rest mass at the top of a barrier of height 0.01 is $\,E_{cr}= 1.2591$. Thus,

 \begin{center}
 $\,u_{cr}=\sqrt{1-\left(\frac{1.2528}{1.2591}\right)^{2}}=0.0999$,
 \end{center}
which is in an excellent agreement with the numerical value mentioned before.

This was for the Q-ball of $\omega=1.9$. What about other values?
In Table 1. we present the calculated critical velocities for several values of $\omega$.

\begin{center}
\begin{tabular}{lllll}
$\omega$ & $\,M_{rest}$ & $\,E_{cr}$ & $u_{cr}$ \\
1.5&  3.2159 & 3.23199  & 0.09938$\sim 0.1$\\
1.6 & 2.6167&2.62985    &  0.0999  $\sim 0.1$\\
1.7 & 2.1892 & 2.2002  & 0.0997   $\sim 0.1$\\
1.8 & 1.7684 & 1.7772  & 0.0993   $\sim 0.1$\\
1.9 & 1.2528 & 1.2591  & 0.0999   $\sim 0.1$ \\

\end{tabular}
\end{center}
\begin{center}
Table 1.
\end{center}

The agreement is clearly very good. This agreement shows that during the scattering very little radiation
is sent out. Moreover, although the energies of the Q-balls depend on their frequency ({\it ie} on their
charge $Q$) the value of the critical velocity is approximately independent of this frequency.
%This independence has been seen also for other heights of the barrier.

We have also looked at barriers of different heights. In particular, when the height is 0.1 we have found that the critical
velocity was around $\sim 0.3$. Our estimates give us again the same value $\sim0.3$ but this time the stability line has increased to $m=2.097$ (see fig. 4) and so the Q-balls with frequency above $\omega=1.73$ 
are unstable. Hence the barrier has divided the parameter space of Q-balls: only those with frequency 
below $\omega=1.73$ are stable and for them the critical velocity is around $\sim 0.3$.

\begin{figure}
\begin{center}
\includegraphics[angle=270, width=10cm]{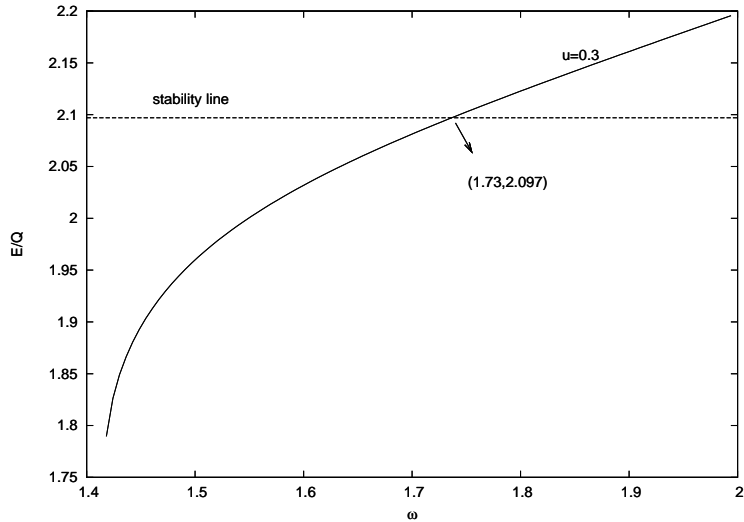}
\caption{ The stability line for a barrier of height =0.1. }
\end{center}
\end{figure}

What happens if we send a Q-ball which is unstable at the barrier?

We have performed such simulations and have found that such a Q-ball moves slower as it tries to climb 
the barrier and its tail gets distorted. When we repeated this experiment with higher barriers
we have found that the Q-ball splits into two or more Q-balls.
 As the barrier is increased further the Q-ball splits into more Q-balls.

This fission starts at the edge of the barrier and the split off Q-balls are reflected 
 while the parent Q-ball slowly climbs the barrier. At the time of splitting the Q-ball is quite
distorted and this sets off the fission process. 
Note that this splitting depends on the frequency of the Q-ball as the Q-balls 
with low $\,Q$ value are more likely to get deformed and so produce more fission during their scattering
on the barrier.

Finally, let us examine the dependence of the repulsive force of the barrier on the frequency {\it ie} the charge
of the Q-balls. To examine this we have placed two Q-balls (of frequencies $\omega=1.5$ and $\omega=1.9$), at rest,  at $\,x=-15$ {\it ie} close to a barrier of height 0.9. In both cases, as expected, the Q-balls moved away from the barrier. Fig. 5 presents the time evolution of their 
positions and it clearly demonstrates that the Q-ball with $\omega=1.5$ has moved slower than the Q-ball with $\omega=1.9$. Hence the forces acting on the Q-ball with $\omega=1.5$ are weaker. This can be easily understood if we consider a Q-ball as a point particle and a barrier as a high wall. When two particles have the same momentum but one is more massive than the other one the momentum conservation implies that the massive one moves backwards slower than the less massive one. As the Q-ball with $\omega=1.5$ is more massive than the one with $\omega=1.9$  it moves away slower from the barrier.

\begin{figure}
\begin{center}
\includegraphics[angle=270, width=10cm]{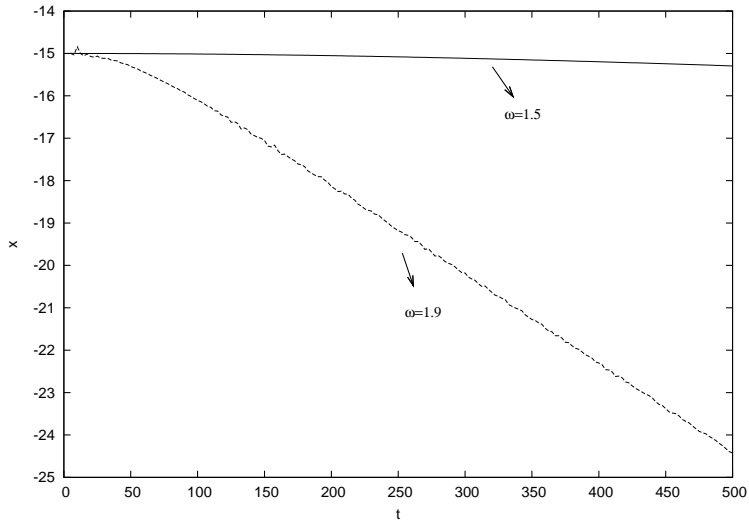}
\caption{  Time dependence of the positions of Q-balls with $\omega=1.5$, $\omega=1.9$ initially at rest and close to a barrier of height 0.9.}
\end{center}
\end{figure}

If we decrease the barrier height, the Q-ball is repelled less. Fig. 6 shows the time 
dependence of the positions of a Q-ball with $\omega=1.9$  which has been repelled by  barriers of height 0.5 and 0.9. We note that the repulsive force decreases with the decrease of the height of the barrier.  

\begin{figure}
\begin{center}
\includegraphics[angle=270, width=10cm]{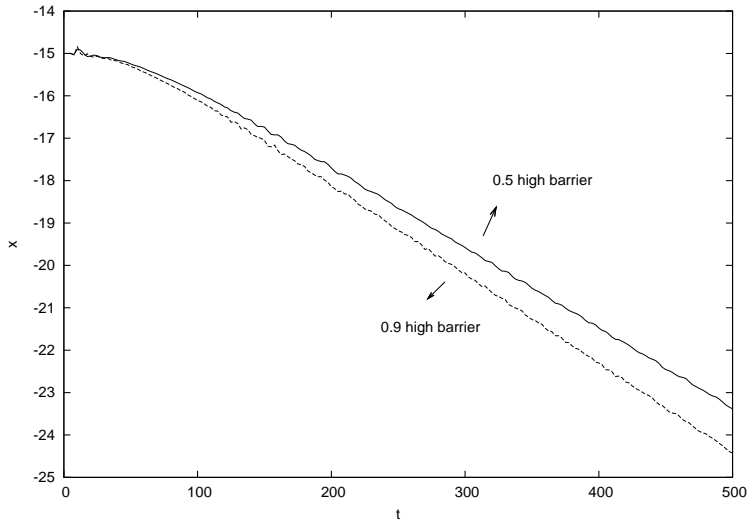}
\caption{ Time dependence of the positions of Q-balls with $\omega=1.9$, initially at rest, for barriers of heights 0.5 and 0.9.}
\end{center}
\end{figure}

 It may appear somewhat surprising, at first sight, that the repulsive force between a barrier and a Q-ball depends on the charge while the critical velocity is independent of it. The Q-balls at the top of the barrier are modified by
the scaling $\lambda\rightarrow \tilde \lambda$ and so we see that only the barrier height determines their critical velocity. We note that the mass of the Q-ball at the top of the barrier is scaled up so that $ \,M_{B}=\sqrt{\tilde\lambda}\,M_{rest}$. Hence, from (32), we see that the critical velocity is also a function only of the barrier height and is given by:
 \begin{equation}
\,u_{cr}=\sqrt{\frac{\lambda_{0}}{1+\lambda_{0}}} .
\end{equation}

Thus, when the height of the barrier is 0.01, the above expression gives for the critical velocity the value of 0.0995. Also, for a barrier of height 0.1 the above expression gives  0.3015. These results are in an excellent agreement with the values calculated and observed by us in our simulations.

\section{\textbf{Scattering of Q-Balls on Holes}}

 The potential obstruction is a hole when $0<\tilde\lambda<1$. Then the potential is scaled down. While for
the barriers the stability line (see fig. 2) was shifted up - for holes it is shifted down and so the stability
becomes more delicate.
Interestingly, when the stability conditions of the Q-balls are satisfied, their scattering on the holes resembles the dynamics of topological solitons. Then a Q-ball far away from obstruction moves in a well defined trajectory and as it encounters the hole it speeds up. If the Q-ball has enough energy it is transmitted giving off a small amount of radiation; otherwise it gets trapped in the hole.

As for holes the potential is scaled down, the corresponding stability line,  {\it ie} the value of the mass, is lowered according to the formula $\,m=2\sqrt{\tilde\lambda}$. Thus, for a hole of depth -0.9 the stability line
is lowered to $m=0.6324$. In such a case all the Q-balls that can be obtained in the initially allowed frequency range will be unstable in the hole even when the Q-ball has zero velocity. This is clearly seen from fig. 2 and this has been confirmed by our numerical simulations. In order to understand how much the Q-ball is affected by a hole as deep as -0.9, we have placed a Q-ball with $ \omega=1.9$, at rest, at $\,x=-14.0$  ({\it ie} close to the hole which is located at $ -10<\,x<10$). As the times evolves the tail of the Q-ball interacts with the hole and generates many smaller Q-balls inside the hole. This can be seen in figure 7.  

\begin{figure}
\begin{center}
\includegraphics[angle=270, width=10cm]{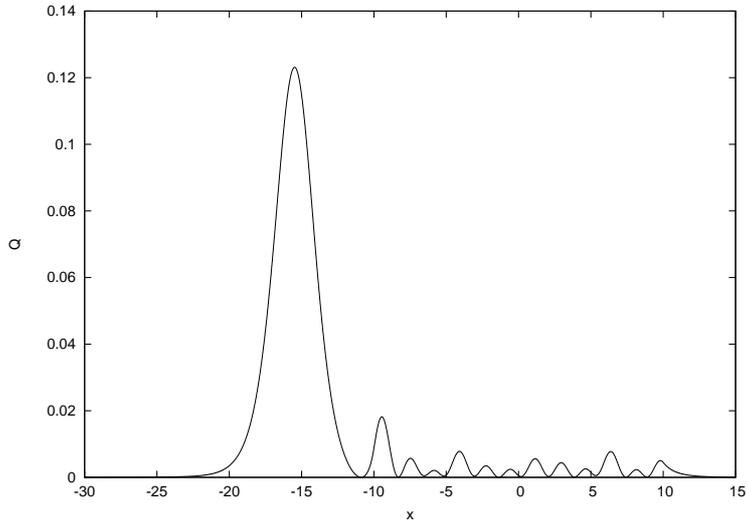}
\caption{The Q-ball $\,Q$ with $\omega=1.9$ creating smaller Q-balls in the hole of depth -0.9.}
\end{center}
\end{figure}

The number of smaller Q-balls which are produced inside the hole keeps on increasing and finally, after some time, settles to a nearly constant number.  We have found that for each Q-ball the number of the small Q-balls produced inside the hole depends  on how deep the hole is. Table 2 summarizes our results. It presents the number of smaller Q-balls created inside the hole for various hole depths when the initial Q-ball had $\omega=1.9$ 
and, as before, was located at $\,x=-14$.
For other values of $\omega$ the results were similar although as for lower values of $\omega$ the charge 
and energy were larger (see fig. 1) and so the process of reaching the equilibrium was slower.

\begin{center}
$\omega=1.9$
\begin{tabular}{llll}
$\lambda_{0}$ & No. of the Q-balls in the Hole \\
 -0.9 & $\qquad$ 12 \\ %repeat
-0.6 & $\qquad$  10  \\
-0.5 & $\qquad$  9   \\
-0.3  & $\qquad$  7   \\
-0.2 & $\qquad$   5
\end{tabular}
\end{center}

\qquad \qquad \qquad \qquad \qquad Table 2.\\

So, in situations where there are some Q-balls created inside the hole, the hole is not passive as was the case with a barrier but rather becomes a charged hole with a very complicated structure especially when there are many small Q-balls that had been produced inside it.

 The force between a Q-ball and a hole is not very simple any more, as was the case for topological solitons. Again, it is the stability condition which, when violated, gives rise to this difference. We have found that when the stability condition is satisfied (and no small Q-balls are generated) the force between a Q-ball and a hole is
 attractive. This corresponds to what was seen for a topological soliton in the presence of holes.
In the Q-ball case, however, we have found that, in general, the interaction can be described as corresponding to:

\begin{itemize}
                                                                                                                                         \item a purely attractive force,
                                                                                                                                                                                                                                                                                 \item  a purely repulsive force,

 \item a mixed force.
                                                                                                                                                                                                                                                                                                                                                                                                                    \end{itemize}

A purely attractive force has been observed when a Q-ball is stable or when a Q-ball may be unstable but has not been distorted too much by the presence of the hole. This can happen when the hole is shallow and the charge of the Q-ball is large. However, when the hole is deep and the Q-ball gets distorted, its tail develops many small Q-balls inside the hole and the force becomes repulsive. This repulsion then prevents the Q-ball
from falling into the hole.

Fig. 8 presents the plots of the time evolution of two unstable Q-balls with frequency $\omega=1.5$ placed, at rest, close to the holes of depth -0.5 and -0.9. We plot the positions of the `parent' Q-balls as the scattering, in each case, generates small Q-balls in the hole. We note that while the shallower hole attracts the Q-ball, the opposite is the case for a deeper hole.
This repulsion is due to the greater distortion of the Q-ball by the hole 
resulting in more small Q-balls in the hole which in turn changes the attraction
into the repulsion.

\begin{figure}
\begin{center}
\includegraphics[angle=270, width=10cm]{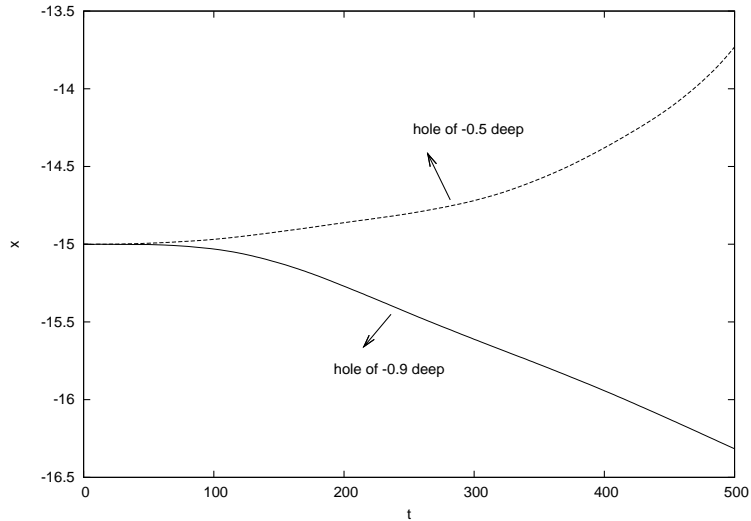}
\caption{Plots of the time dependence of the position of a Q-ball with $\omega=1.5$, $\,u=0$, placed near holes of depth -0.5 and -0.9} 
\end{center}
\end{figure}

In some of our simulations we have also observed that an unstable Q-ball could be,
initially,  attracted by the hole and then repelled by it. This occurs when the hole is not very deep and the tail of the Q-ball requires some time to generate the smaller Q-balls inside the hole. In this case, before generating these small Q-balls, the `parent' Q-ball  behaves as a normal soliton, {\it ie} is attracted by the hole. However, when the tail has 
already generated  sufficient number of little Q-balls in  the hole the repulsive force  begins to dominate. Fig. 9  describes such a case. It presents the plot of the position
of a `parent' Q-ball with $\omega=1.9$ when the hole is of depth -0.125.

\begin{figure}
\begin{center}
\includegraphics[angle=270, width=10cm]{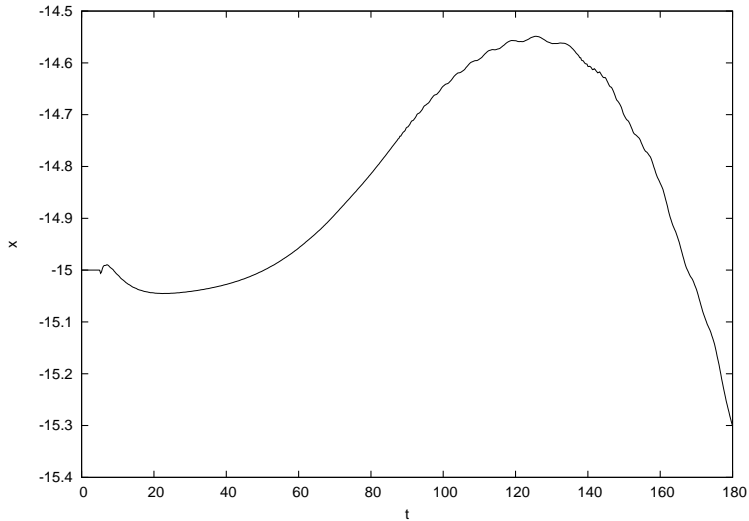}
\caption{The dependence of the position of a `parent' Q-ball with $\omega=1.9$, $\,u=0$, 
placed at rest close to a hole of depth -0.125.}
\end{center}
\end{figure}

The above mentioned results suggest that one should try to explain them in terms
of the forces between Q-balls. Not much is known about such forces. In [12]
some results are given; in particular, they show that two Q-balls with the same frequency
$\omega$ attract each other if their relative phase $\theta$  is $0$ and repel if
$\theta=\pi$. In our case, the frequencies are different and so the forces (between
the `parent' Q-ball and it `offspring') become 
effectively time dependent. 

%It has been studied in Ref. [10], the force between two well-separated Q-balls on a small interval far from both Q-balls, having two internal frequencies $\omega_{1}$ and $\omega_{2}$ with $\theta$ the relative phase at t=0. It has been found when $\omega_{1}=\omega_{2}$ that the force will be attractive when the two are in phase, $\theta=0$ and repulsive when they are out-of-phase, $\theta$=$\pi$. However, when $\omega_{1}\neq\omega_{2}$ the net force averages to zero on that small interval. The two Q-balls with distinct frequencies have a breather-like motion. But non-of these results apply to our case. The hole when the parent Q-ball creates some little Q-balls inside the hole has a very complicated structure with many distinct frequencies and phase differences. 

Given this situation we have performed further numerical studies. Thus we have placed 
Q-balls of frequency $\omega=1.5$ and $1.9$ far away from the hole of depth -0.9 
and sent them with velocities 0.1 towards the hole. Both Q-balls moved towards the hole and when 
their tails reached the hole they started generating small Q-balls inside it. 
The small Q-balls inside the hole had various frequencies and so interacted with each other
and with their `parents'. These interactions amongst the `baby' Q-balls let them 
oscillate inside the hole and ultimately led to the repulsion of the `parent' Q-ball.
In fig. 10 we present the plots of the positions of the `parents'.  We note that the Q-ball
with $\omega=1.5$ gets nearer to the hole but ultimately is reflected with a higher
velocity than the Q-ball of $\omega=1.9$. This can be partly understood by noting that
although both Q-balls are unstable the one with with $\omega=1.5$ has a higher charge and hence requires a longer time for its tail to generate the Q-balls inside the hole. But once they are there their repulsion is stronger.

%Now, if a Q-ball is located far away from a hole of -0.9 deep and it has been given a small velocity of 0.1, the Q-ball will move toward the hole. As the Q-ball is getting closer to the hole, the tail of the Q-ball will develop smaller Q-balls inside the hole. Because the hole contains many  smaller Q-balls each might be rotating with different internal frequency, the parent Q-ball in case of low velocity will be reflected back,{\ it i.e} repelled by the hole. The smaller Q-balls will oscillate back and forth inside the hole and will continually fragmented at the edge of the hole, though keeping the number of the smaller Q-balls  nearly fixed inside the hole.  The parent Q-ball which has been reflected back from the hole will partially absorbed at the boundary. 

\begin{figure}
\begin{center}
\includegraphics[angle=270, width=10cm]{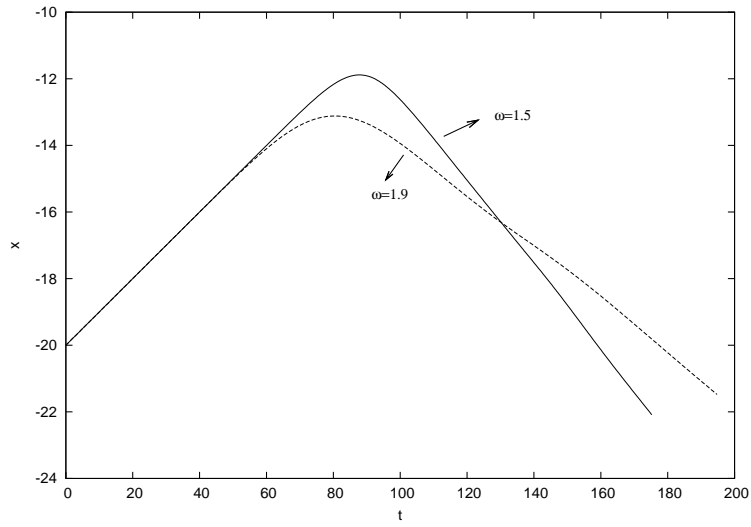}
\caption{The time dependence of the positions of Q-balls with $\omega=1.5$ and $\omega=1.9$, initially sent with velocity $u=0.1$ towards the hole of depth -0.9.}
\end{center}
\end{figure}

If we increase the velocity of the Q-balls, their tails develop a smaller number of 
`baby' Q-balls as if such Q-balls were more stable. In fact, this is not the case; a Q-ball moving with a higher speed is less stable but its interaction time with the hole
gets shorter. Thus a Q-ball with $\omega=1.9$, sent with velocity 0.95,
 has too little time to generate many `baby' Q-balls inside it and gets through the hole
 without much distortion.

If we consider a shallow hole the stable Q-ball behaves very much like a topological soliton. So it moves along a well defined trajectory and as it approaches the hole it speeds up. If the Q-ball has enough energy it comes out at the other end of the hole having
generated  very tiny radiation, otherwise it gets trapped in the hole. Fig. 11
presents the plot of the time dependence of the position of  a Q-ball with $\omega=1.5$ sent 
towards the hole of depth -0.1 with velocity just below its critical value for transmission 
 (in this case $u=0.023$). 

\begin{figure}
\begin{center}
\includegraphics[angle=270, width=10cm]{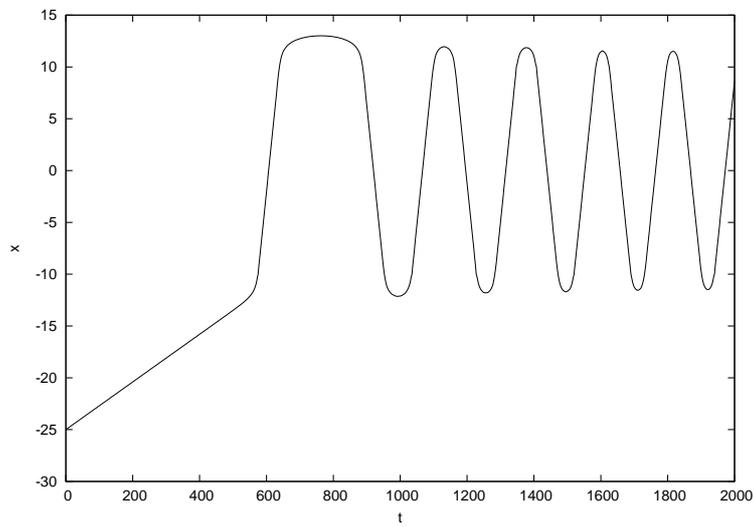}
\caption{Time dependence of the position of  a Q-ball with $\omega=1.5$ sent 
towards the hole of depth -0.1 with velocity $\,u=0.023$.}
\end{center}
\end{figure}

\begin{figure}
\begin{center}
\includegraphics[angle=270, width=10cm]{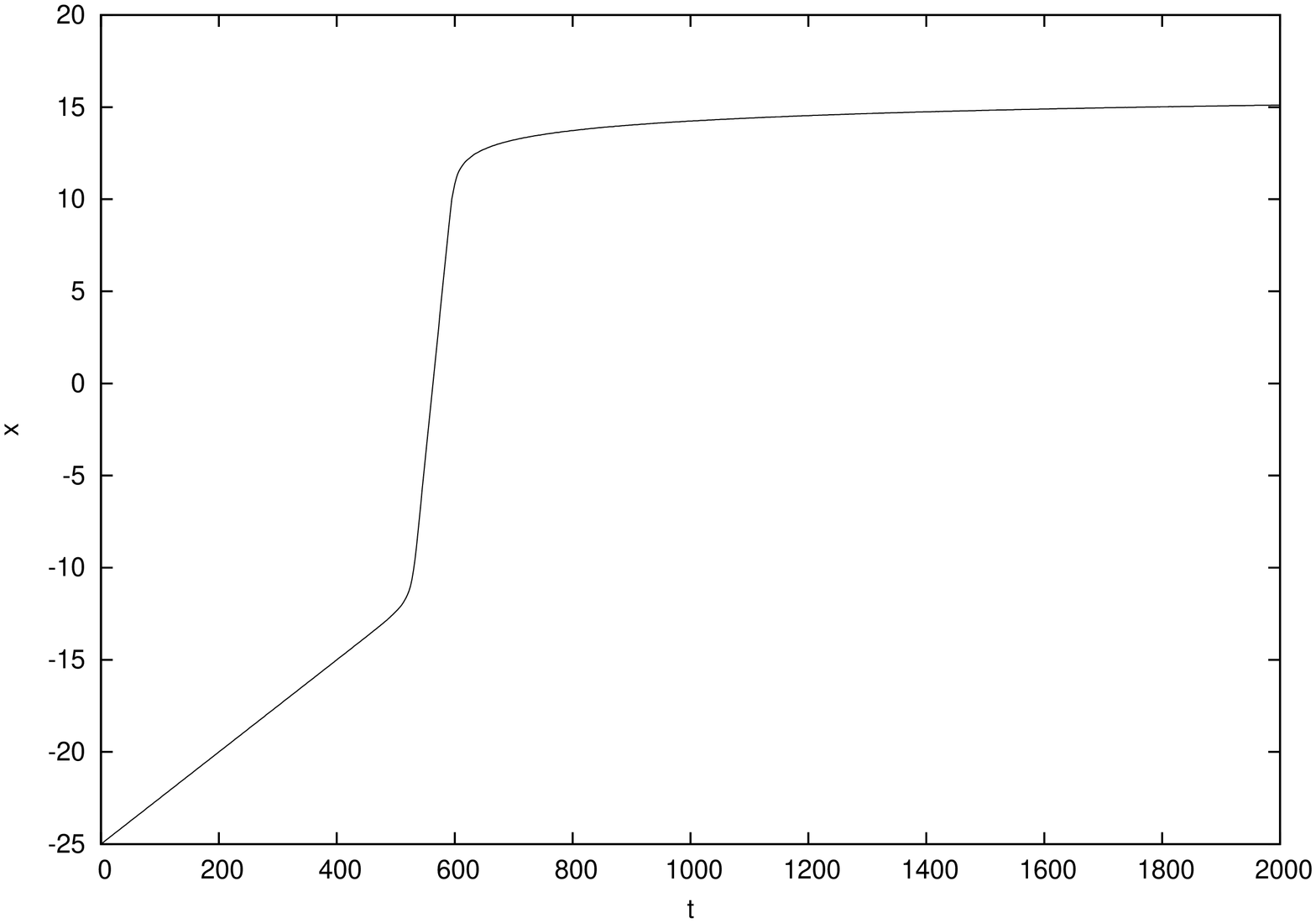}
\caption{Time dependence of the position of  a Q-ball with $\omega=1.5$ sent 
towards the hole of depth -0.1 with velocity $\,u=0.025$}
\end{center}
\end{figure}

For a velocity slightly larger we observed the transmission - shown in fig. 12. Hence 
the critical velocity for transmission is close to $0.024$.
Next we looked at the dependence of this critical velocity on the frequency 
of the Q-ball ({\it ie} its $\omega$). Our results are presented in Table 3.

%For a hole of -0.1 deep, we have found numerically the critical velocities for the possible stable Q-ball solutions in the allowed limit. We have selected a hole that leaves the Q-ball stable during the scattering running after not an easy task of finding the critical velocities for each solution. Table 4 summarizes our results for the critical velocities found for a hole of -0.1 deep for various Q-ball solutions.

% EITHER GRAPH OR TABLE

 \begin{center}
\begin{tabular}{llll}
$\omega$ & $\,u_{cr}$ \\
 1.5 & $\sim$ 0.025 \\
 1.55 & $\sim$0.04385    \\
 1.6 & $\sim$0.0445    \\
 1.65  & $\sim $0.072 \\
    
\end{tabular}
\end{center}
\begin{center}
Table 3.
\end{center}

As can be seen from the table the critical velocities increase as the charge $\,Q$ decreases.
The observed irregularities prevent us from drawing too detailed conclusions. Clearly, the
exact values depend on the details of the interaction with the hole (the same phenomenon
was observed for topological solitons - see \cite{extra}).

However, some information can be drawn from the fact that for a hole of depth -0.1 the
stability line is lowered from $m=2$ down to $m=1.89$. As can be seen from fig. 13 this affects
the critical velocities for $\omega=1.7$ and larger. Looking at fig. 13 we note that,
as the velocity increases, fewer Q-balls (with higher $\omega$) are stable.
Thus a Q-ball with $\omega=1.65$  has as its critical velocity $u_{cr}\sim 0.07$ which is just below the intersection point. However, for a Q-ball of frequency $1.7$ and above 
we would expect its critical velocity to be greater than the one for $\omega=1.65$ and hence 
such Q-balls will not be stable. This is, in fact, what has been observed in our
numerical simulations. So, all velocities that lie above the intersection point would make the Q-balls unstable. Such Q-balls would generate `baby' Q-balls and would not be transmitted. Hence, in such cases we would not have critical velocities. To be specific 
the critical velocity for a Q-ball with $\omega=1.65$ is $\,u\sim0.075$ and so at this velocity the intersection point is at $\omega=1.664$. As $\omega=1.65$ is below the stability line the Q-ball is stable and it has a critical velocity.

\begin{figure}
\begin{center}
\includegraphics[angle=270, width=10cm]{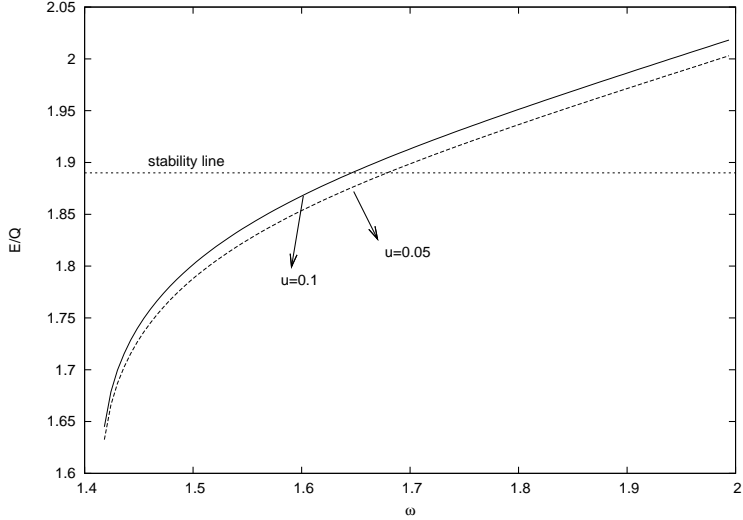}
\caption{The stability line  for a hole of depth -0.1.}
\end{center}
\end{figure}

%XXXXXXXXXXXXXXXXXXXXXXXXXXXXXXXXXXXXXXXXXXXXXXXXXXXXXXXXXXXXXXXXXXXXXXXXXXXXXXXXXXXXXXXXXXXXXXXXXXXXXXXXXXXXXXXXXXXXXXXXXXXXx

\section{\textbf{Q-ball Scattering in (2+1) Dimensions}}

 Next we consider the scattering of Q-balls in two spatial dimensions. As before 
 we consider their behaviour in the presence of various 
 potential obstructions.  %The Q-ball scattering in $\left(2+1\right) $ dimensions has given further confirmation of the particle nature of such extended structure in the limit when the Q-ball tail far enough from the Q-ball body is involving in the interactions. The Q-ball scattering on obstructions have shown great resemblance to some known scatterings in atomic and electromagnetism. Because the Q-ball has a continuous charge the Q-ball behaves like a charged point particle. By resemblance we mean that the parameters that affect the Q-ball scattering are the same parameters that play the role in controlling these known scatterings in physics.

We return to eq. (1) and note that this time the index $\mu$ takes the values $\mu=0,1,2$. The field equation becomes

\begin{equation}
 \ddot\Phi-\nabla^{2}\Phi+2\Phi\tilde\lambda\left(2-4\vert\Phi\vert^{2}+3\vert\Phi\vert^{4}\right)=0,
\end{equation}
 where $\nabla^{2}=\partial_{\,x}^{2}+\partial_{\,y}^{2}$. The Q-ball ansatz becomes

\begin{equation}
 \Phi\left(\,r,\,t\right)=\,f\left(\,r\right)\,exp\left(\,i\omega\,t\right),
\end{equation}
where $\,r=\sqrt{\,x^{2}+\,y^{2}}$. The equation $\left(10\right)$  for the profile 
field $f$ now becomes

\begin{equation}
\frac{\,d^{2}f}{\,dr^{2}}+\frac{1}{r}\frac{\,df}{\,dr}+\omega^{2}\,f-2\tilde \lambda\,f\left(2-4\,f^{2}+3\,f^5\right)=0.
%\,\tilde U^{\prime}\left(\,f\right)=0.
\end{equation}

Here, as before,  $\tilde\lambda=1+\lambda_{0}$.

As before the profile functions $f$ were determined numerically. Fig. 14 
presents the plots  of these profiles for various values of $\omega $ in the allowed range, {\it ie} $\sqrt{2}<\omega<2$. Having found the static Q-balls we have boosted 
them along the $x$-axis by performing the appropriate Lorentz transformation, ${\it ie}$:
\begin{equation}
 \,x^{\prime}=\gamma\left(\,x-\,x_{0}-\,u\,t\right),\\
\qquad  \,y^{\prime}=\,y,\\
 \qquad \,t^{\prime}=\gamma\left(t-\,u\left(\,x-\,x_{0}\right)\right),
\end{equation}
where, as usual, $\gamma=\frac{1}{\sqrt{1-\,u^{2}}}$.  
This allows us to study the interactions of Q-balls as they, initially, move along the 
$x$ axis.

We have considered various potential obstructions. Most of our simulations have involved 
circular obstructions of radius $R=5$. They were centered at $(x_0,y_0)$. Looking 
at the profiles in fig. 14 we note that for some $\omega$'s the Q-ball would not easily fit into the hole.  However, in this work we have been primarily interested in the main aspects
of the dynamics external to the holes - so for this such holes were sufficient. 
In our simulations we used a grid containing $300^{2}$ points with $\,dx=0.1,\,dy=0.1$ and $\,dt=0.02$. This, the size of the grid and the size of the hole allowed us to study 
the dependence on the impact parameter of the scattering without having to worry about the edge effects.

%The Q-ball  moves along the x-axis and interacts with an obstruction,{\it i.e.} a barrier or a hole which are located along the y-axis,{\it i.e.}$\left(\,x_{0},\\,y_{0}\right)=\left(0,\,y_{0}\right)$. Where $\,x_{0},$ and $\,y_{0}$ are the positions of the obstruction on the x-and y-axes with $\,x_{0}=0$. The obstruction is made circular with a radius, $\vert \,R \vert = 5$. The distance from the centre of this circle to x-axis represents the impact parameter, $\,y_{0}$, of the scattering. The width of the circle is not necessarily enough to accommodate the Q-ball since we are not interested in a direct collision with the Q-ball, but it should be enough to reveal the dynamics of the scattering. Our choice of the width is enough to do that. We have used a grid containing $300^{2}$ points with $\,dx=0.1,\,dy=0.1$ and $\,dt=0.02$. 

\begin{figure}
\begin{center}
\includegraphics[angle=270, width=14cm]{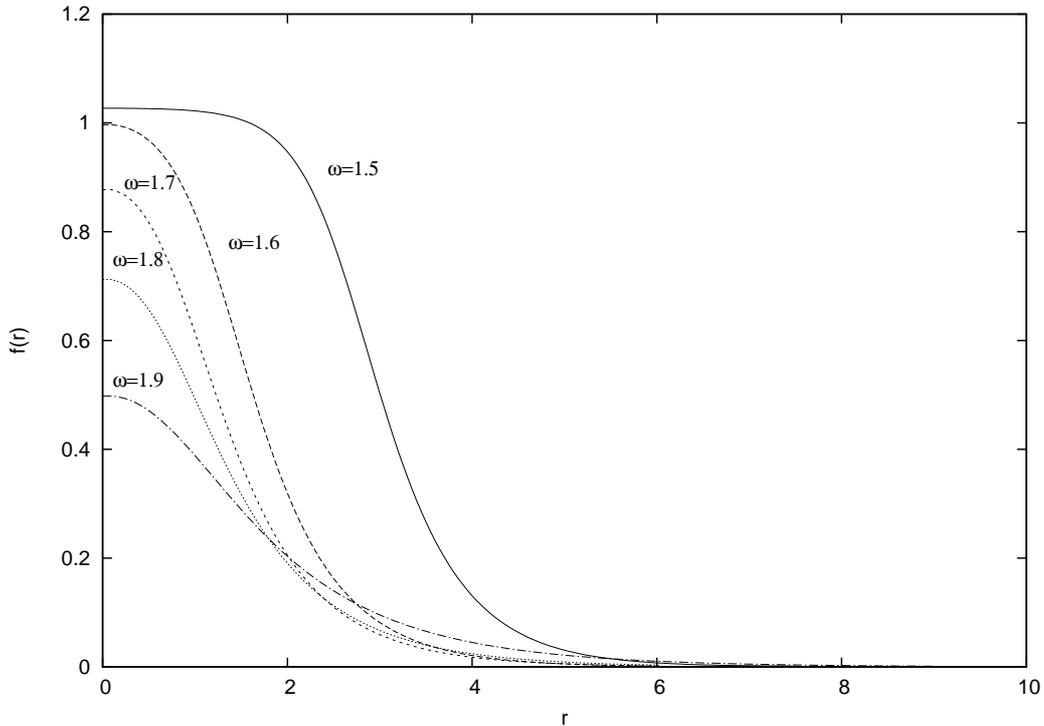}
\caption{The numerically determined profiles functions $f$ for $\omega=1.5,1.6,1.7,1.8,1.9$}
\end{center}
\end{figure}

\subsection{\textbf{Q-ball Scattering on Barriers in $\left(2+1\right)$ Dimensions}}

 The scattering of Q-balls on barriers in $\left(1+1\right)$ dimensions, discussed
 in the previous sections,  has given a clear confirmation of a particle-like nature of 
 Q-balls. This is further supported by our results on the scattering in $(2+1)$ dimensions.
 As we have seen, the basic forces are repulsive and for each barrier there is a critical
 velocity above which the Q-ball gets transmitted over the barrier.
 In $\left(2+1\right)$ dimensions, this force is also repulsive but this time the 
 behaviour depends also on the impact parameter of the scattering.

  To study this we have looked at the scattering of a Q-ball on a series of potential
  barriers whose position was varied along the $y$ axis (so that the scattering took
  places at different impact parameters).
  In fig. 15 we present plots of the trajectories of Q-balls with $\omega=1.75$ 
  sent with velocity 0.1 at barriers of height 0.9 placed at ($0,y_0$) for $y_0=0,3., ...10$.

 %The dynamics of Q-ball scattering from a barrier located along the y-axis resembles in the case of a very high barrier Rutherford scattering of an alpha particle from a very heavy nucleus. The scattering depends on the impact parameter,{\it i.e}, position of the barrier along the y-axis and on the velocity or the energy of the Q-ball. Figure 14, represents a Q-ball scattering on a barrier of 0.9 high when the Q-ball with $\omega=1.75$ was sent with a velocity of 0.1 and the impact parameter has been changed along the y-axis. 

\begin{figure}
\begin{center}
\includegraphics[angle=270, width=14cm]{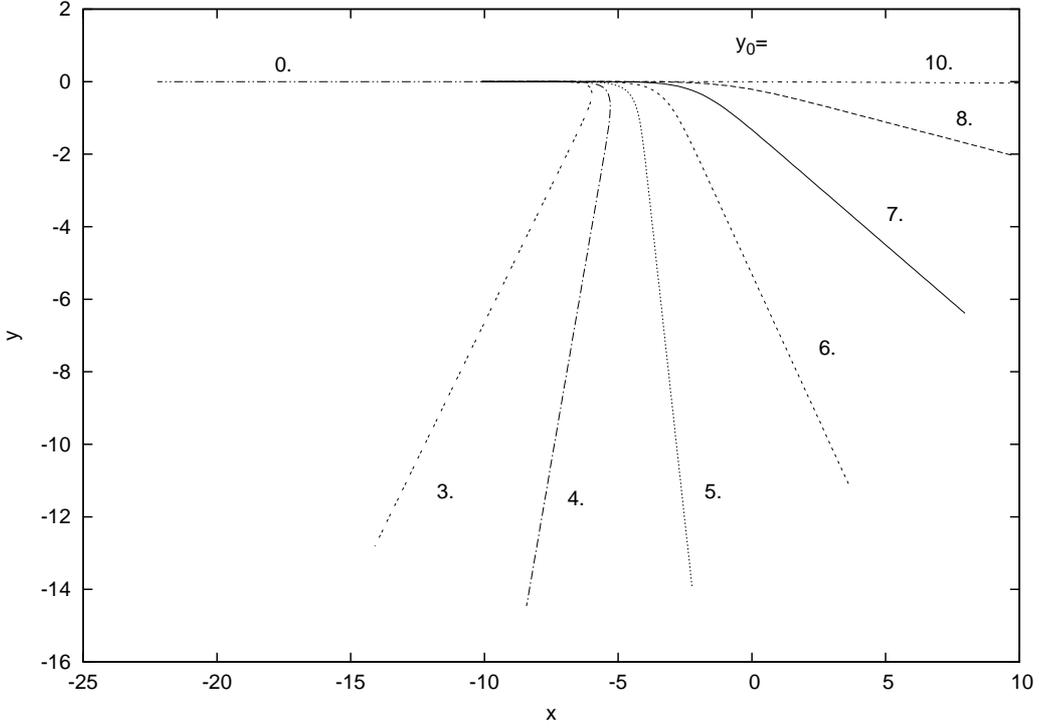}
\caption{Q-ball Scattering from barriers placed at various positions along the y-axis;  $\omega=1.75$}
\end{center}
\end{figure}

The plots also indicate that the Q-ball is deflected by the barrier 
before it reaches it with its centre. Hence the deflection is due to the
interaction of the Q-ball tail.  The deflection angle depends on the impact parameter.
As the hole has radius 5 and the size of the Q-ball is essentially 5
we expect very little deflection when $y_0\sim10$. This is, in fact, what is
seen from the plots in fig. 15. As the impact parameter decreases the angle of
reflection increases and at zero impact parameter the Q-ball is fully reflected.
Table 4 presents the deflection angles seen in the scatterings.

\begin{center}
\begin{tabular}{llll}
$\,y_{0}$ & $\theta^{\circ}$ \\
 10.0 &   0.5189  \\
  8.0  & 8.555  \\
  7.0  &  31.955\\
  6.0  &   57.84 \\
  5.0 &  81.159   \\
  4.0 &   103.63 \\
  3.0 &  123.68 \\
  0.0 &   180
  
\end{tabular}
\end{center}
\begin{center}
Table 4.
\end{center}
%\caption{The deflection angles of a Q-ball with $\omega=1.75$ scattered from barriers placed at various positions along $y-$axis. }

 These angles have been calculated numerically by using
\begin{equation}
 \theta=\,tan^{-1}\left(\frac{\,u_{y}}{\,u_{x}}\right),
\end{equation}
where $\,u_{x}$ and $\,u_{y}$ are the velocities along the $x$- and $y$- axes, respectively.

How do these results depend on the height of the barrier or the frequency of the Q-ball?
Or on the velocity of the Q-ball?

 We have performed further studies and have found that when a Q-ball is sent at higher speed, the deflection angle is smaller. This is clearly due to the fact that at higher
 velocity the Q-ball has more power to overcome the barrier and so gets nearer
 to the centre of the barrier. What about the dependence on the height
 of the barrier?   Fig. 16 presents the trajectories of Q-balls of frequency $\omega=1.75$
 sent from $y_0=7$ towards the barriers of height 0.5 and 0.9. We note that the deflection 
 angle is smaller for the smaller barrier; again, in this case the Q-ball gets nearer
 to the centre of the barrier.

\begin{figure}
\begin{center}
\includegraphics[angle=270, width=14cm]{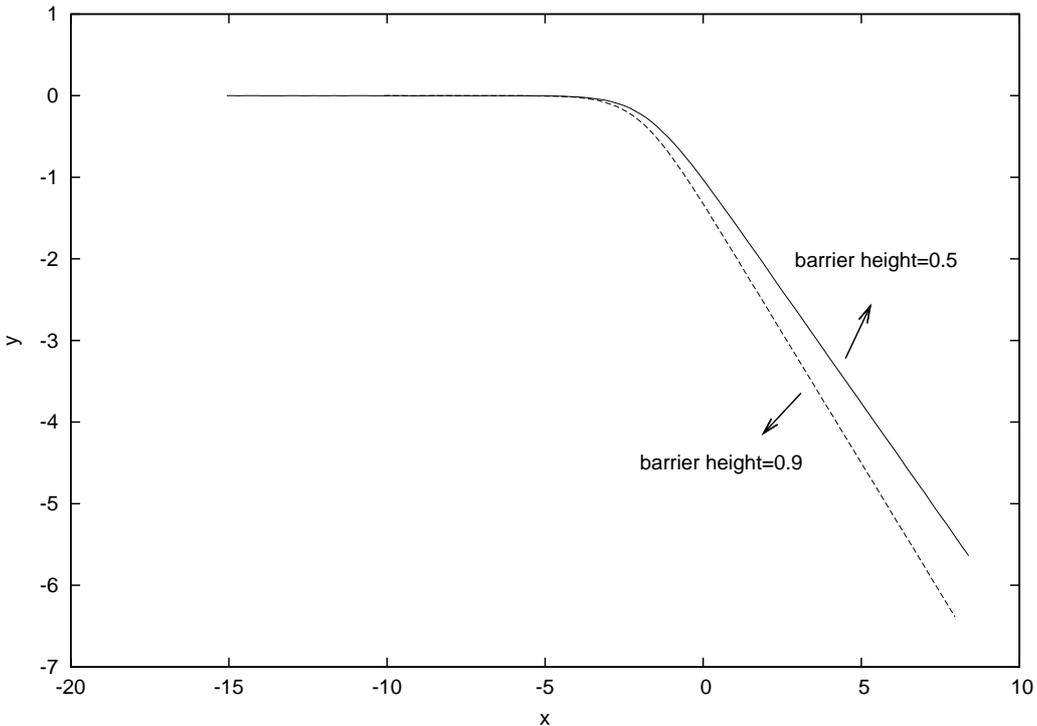}
\caption{Trajectories of Q-balls with $\omega=1.75$ sent
from $y_0=7.0$ with velocity $u=0.1$ towards barriers of height 0.9 and 0.5.}
\end{center}
\end{figure}

Furthermore, we have also found that the deflection angle increases with the decrease
of the Q-ball charge ({\it ie} $\omega$). This is shown in fig. 17 in which we present
plot of trajectories of two Q-balls (with $\omega=1.75$ and 1.9) sent from $y_0=0.0$  with velocity 0.1 towards a barrier of height 0.9 located at $\left(\,x_{0},\,y_{0}\right)=\left(0,7.0\right)$.  The deflection angles for the Q-balls with $\omega=1.75$ and $\omega=1.9$ are found to be 31.955 and 40.167, respectively.

\begin{figure}
\begin{center}
\includegraphics[angle=270, width=14cm]{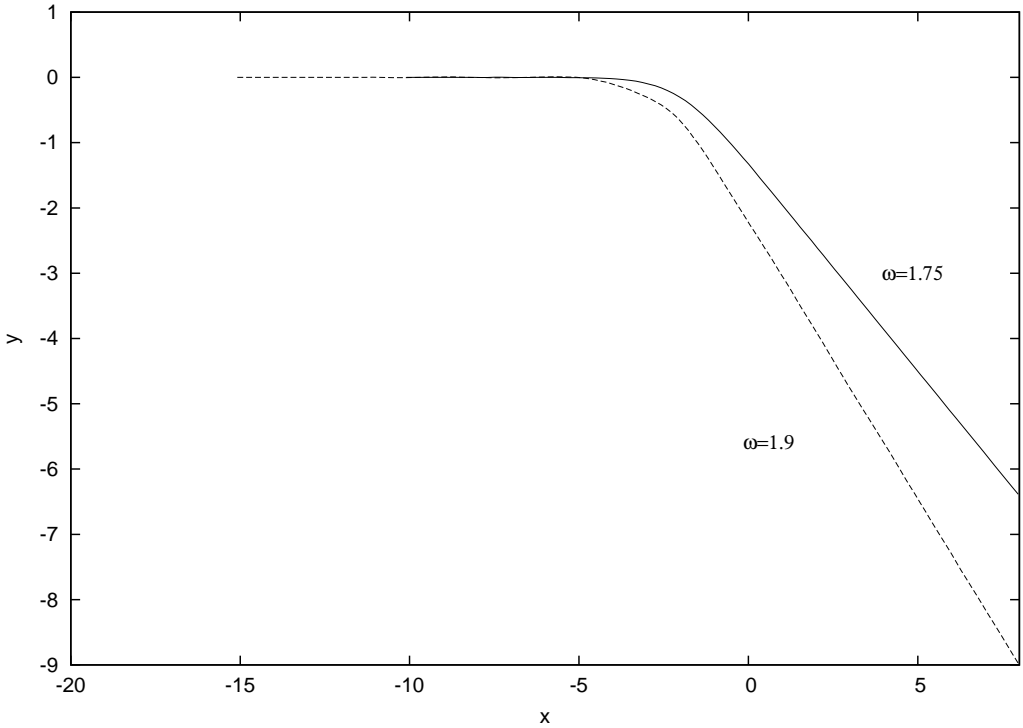}
\caption{Trajectories of Q-balls with $\omega=1.75$ and $\omega=1.9$ sent from $y_0=0.0$ with $u=0.1$ towards a barrier of height 0.9 located at $0.0,7.0$}
\end{center}
\end{figure}

%We have studied numerically the internal frequency when a Q-ball solution has interacted with a barrier. We have used the equation given below to calculate omega. In general, we have found that the internal frequency of a Q-ball decreases at the scattering region.

%\begin{equation}
 %\frac{\vert\frac{\,d\Phi}{\,dt}\vert-\vert\left(\,u_{x}\gamma_{x}\frac{\,d\Phi}{\,dx}+\,u_{y}\gamma_{y}\frac{\,d\Phi}{\,dy}\right)\vert}{\gamma_{x}\vert\Phi\vert}<\vert\omega\vert< \frac{\vert\frac{\,d\Phi}{\,dt}\vert+\vert\left(\,u_{x}\gamma_{x}\frac{\,d\Phi}{\,dx}+\,u_{y}\gamma_{y}\frac{\,d\Phi}{\,dy}\right)\vert}{\gamma_{x}\vert\Phi\vert}
%\end{equation}

%where the subscript x and y refer to the quantities along the x- and y- axes respectively. By considering the maximum value of the field and the derivative of the field with respect to the time and the coordinates we were able to have numerical calculation of $\omega$ within a certain range.  

%%%%   I am not sure what we have demonstrated here!

\subsection{\textbf{Q-ball Scattering on Holes in $\left(2+1\right)$ Dimensions}}

We have also studied the Q-ball scattering on holes. Like in the ($1+1$) dimensional
case the whole process is more complicated than the scattering on barriers.
As in the $(1+1)$ dimensional case the complications arise from the generation 
of `baby' Q-balls in the hole and their repulsion. Also the smaller the velocity 
the more time the `parent' Q-ball has for generating these `baby' Q-balls.

This, coupled with the dependence on the impact parameter, makes the whole process
very complicated. So here we present our preliminary results - leaving more detailed
investigations for the future. However, these preliminary results are interesting 
enough to report them here.

 %The Q-balls scattering from a hole in two spatial dimensions is very interesting. Our simulations has revealed that a Q-ball moving a long the x-axis when interacts with a hole located along the y-axis  behaves in some respect to an electromagnetic scattering of two point charges or a scattering of a point charge from a collective of many point charges carrying opposite charges. The Q-ball scattering, like the electromagnetic scattering, depends on initial velocity of the Q-ball and on the impact parameter; the distance of the potential hole from the x-axis. 
%Our numerical simulations have shown that the smaller the velocity of the Q-ball the more effective the scattering is. When a Q-ball is moving with small velocity, the interaction between the hole and the Q-ball takes a longer time and hence the effect  will be more pronounced. On the other hand when a Q-ball comes across a hole with a high velocity the interaction time is small and the effect is negligible,{\it i.e} the trajectory gets marginally affected. 

First of all let us note that, like in the ($1+1)$ dimensional case,
 the overall force between the Q-ball and the hole can be repulsive, attractive or a mixture of these two and this depends on the velocity of the Q-ball,  its charge and the
 position and the depth of the hole. This dependence is mainly due to the conditions of the stability of the Q-ball. When the Q-balls, during the process of the scattering, remain stable
  the force between them and the hole is purely attractive and the process resembles 
  the scattering on the barriers except that this time the deflection is towards the hole. 
  
However, the behaviour depends on the initial velocity; the faster the Q-balls move the less
they are deflected by the hole. 
Fig. 18 presents the trajectories of 3 Q-balls (of $\omega=1.6$)  sent with velocities $\,u=0.1, 0.05$ and $0.01$ when the hole (of depth -0.9) was located at $(0.0,\,11.5)$. The plots show very clearly that the force between the Q-ball and the hole is attractive
but the deflection  decreases with the increase in the speed of the Q-ball.

\begin{figure}
\begin{center}
\includegraphics[angle=270, width=14cm]{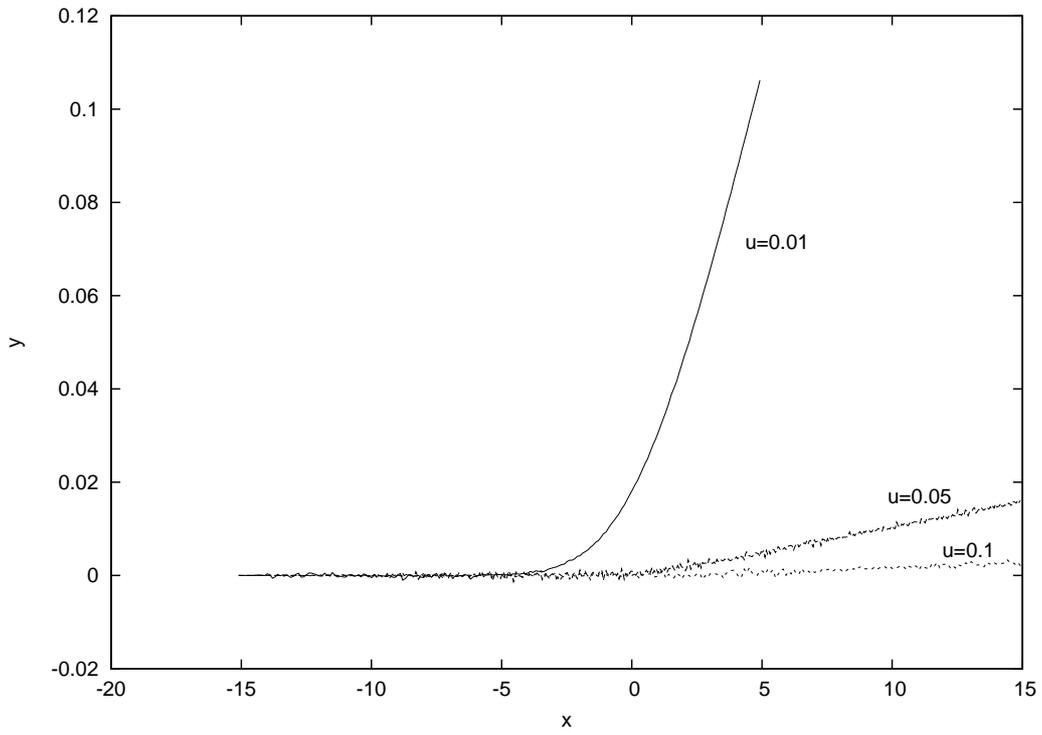}
\caption{Trajectories of Q-balls ($\omega=1.6)$ sent with various incoming velocities for a hole of  depth -0.9 located at $\,y_{0}=11.5$.}
\end{center}
\end{figure}

Next we report the results of our study of the dependence on the impact parameter.
Fig. 19 presents the trajectories of the Q-ball ($\omega=1.6)$  sent with velocity 
 $\,u=0.1$ at various impact parameters, {\it ie} $\,y_{0}=10, 10.5$, and $11$. As expected,
 we see that the deflection increases as we decrease the impact parameter (when the Q-balls
 are sent with the same velocity). 

\begin{figure}
\begin{center}
\includegraphics[angle=270, width=14cm]{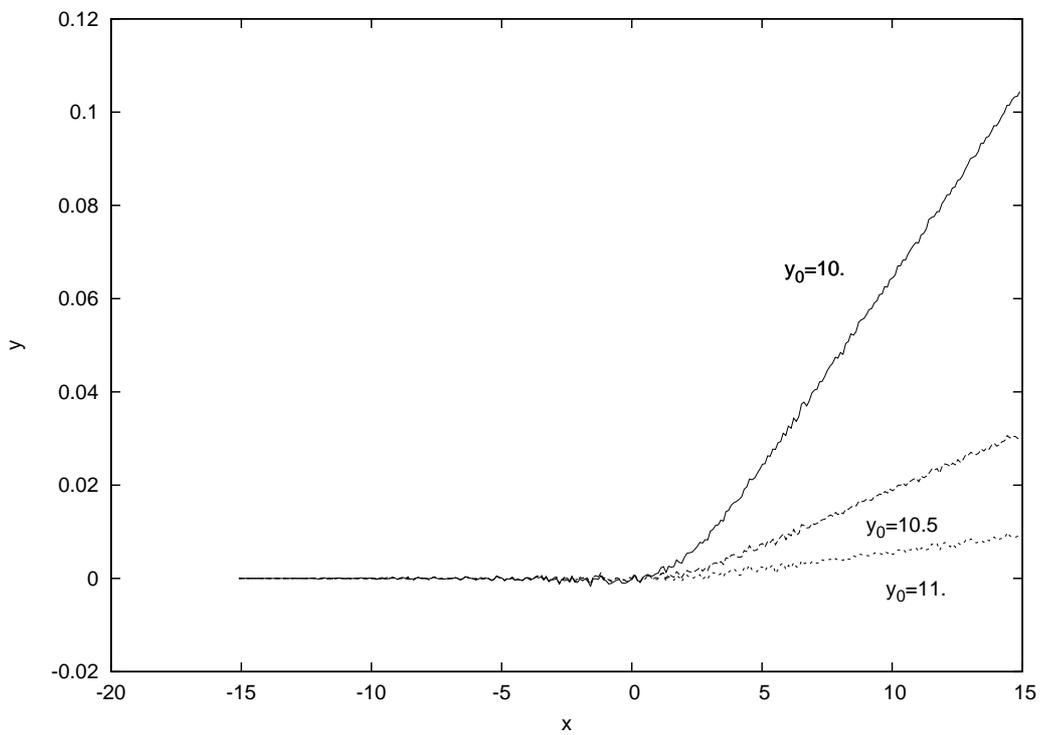}
\caption{Trajectories of a Q-ball  sent with velocity $(u=0.1)$ for a hole of depth -0.9 located at various impact parameters.}
\end{center}
\end{figure}

As we decrease the impact parameter $\,y_{0}$ further,  bringing the hole closer to the $\,x$-axis, the net interaction between the Q-ball and the hole ceases being purely attractive  and becomes partially repulsive and  and partially attractive.
  This  results in a complicated behaviour which is due to what we have seen in the one-dimensional scattering: the hole is no longer a passive object but rather a charged object (due to the `baby Q-balls' that have been generated in the hole) and that affects the interaction.  Fig. 20  presents the trajectory of the `parent' Q-ball (of $\omega=1.6$) sent with  velocity 0.2 towards a hole located at $\,y_{0}=8.0$. The figure shows very clearly that initially the scattering is repulsive and later becomes attractive.
Note the small scale of the $y$-axis which shows that the motion in the $\,y$ direction is very small.
We have also checked that if we increase the velocity the repulsive behaviour disappears. 
Hence the behaviour is very sensitive to the values of various parameters (velocity or impact
parameter).

\begin{figure}
\begin{center}
\includegraphics[angle=270, width=14cm]{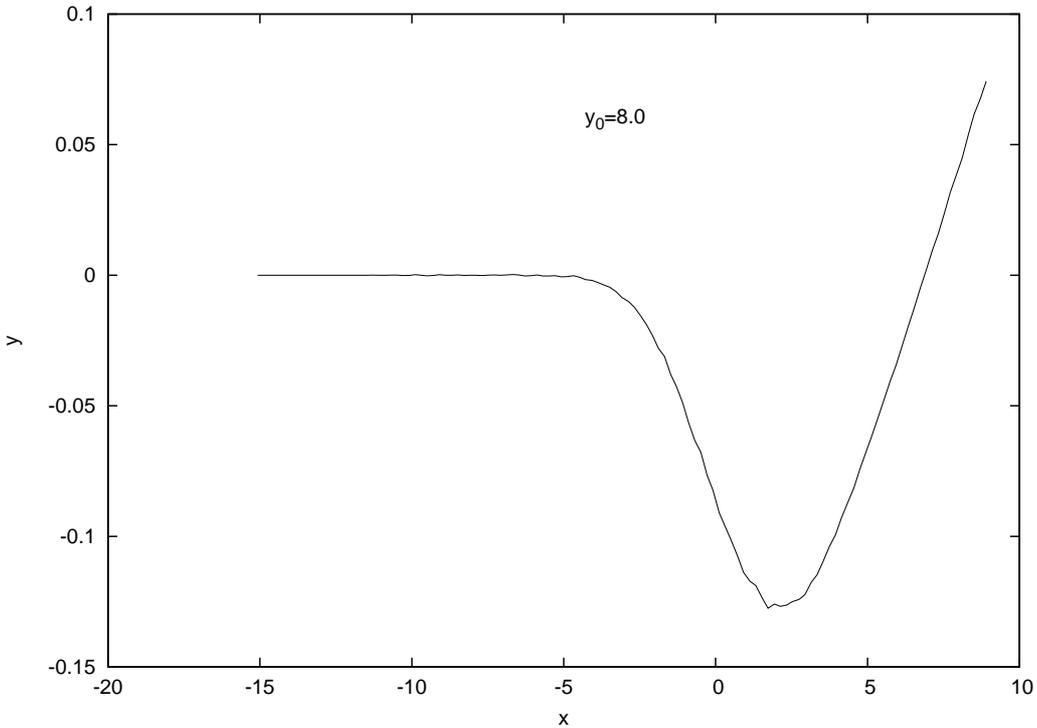}
\caption{Trajectory of a Q-ball (of $\omega=1.6$) sent with velocity $u=0.2$ towards a hole of depth -0.9  located at $\,y_{0}=8.0$.}
\end{center}
\end{figure}

Furthermore, when a Q-ball is sent with a small velocity, it can interact with the hole earlier. At this stage the force is attractive and the Q-ball has not yet
had the time to generate many `baby' Q-balls in the hole. Thus the repulsive force is smaller or even disappears altogether and a purely attractive force is observed. Fig. 21 
presents the trajectories of 2  Q-balls (with $\omega=1.6$) sent with  velocities $\,u=0.05,$ $\,u=0.2$ towards a hole located at $\,y_{0}=8.0$. From the figure one  sees very clearly that the repulsive force seen in the case of $\,u=0.2$  disappears when the Q-ball velocity is decreased to $\,u=0.05$ and that, in this latter case, the  overall force is attractive. 

\begin{figure}
\begin{center}
\includegraphics[angle=270, width=14cm]{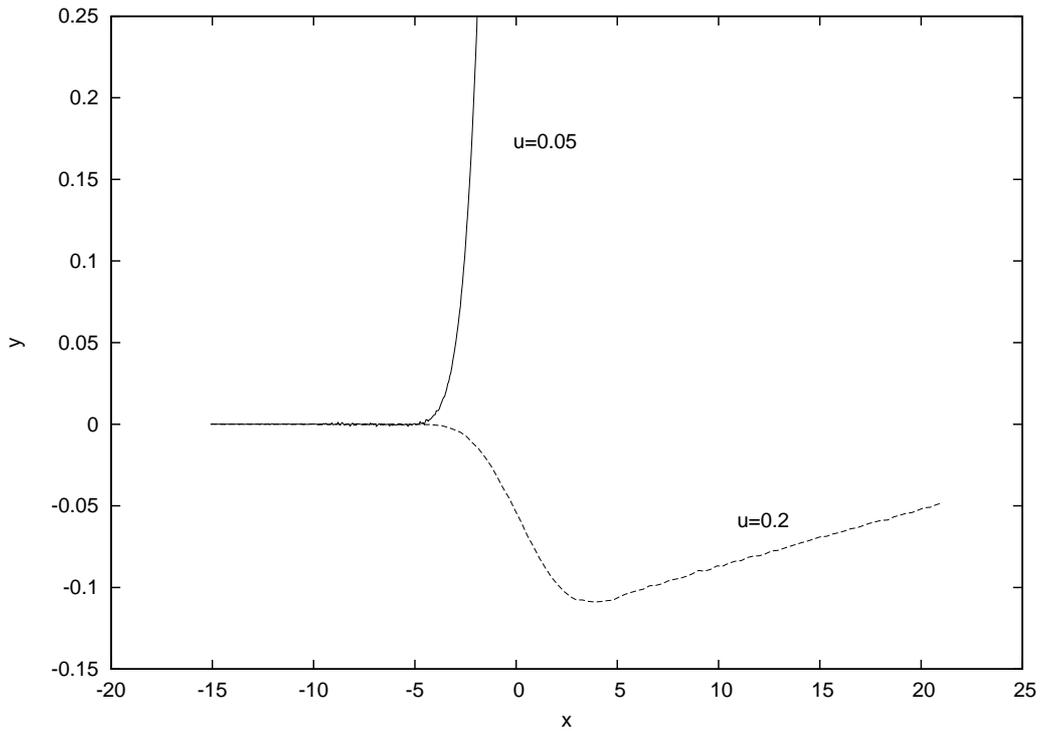}
\caption{ Trajectories of 2 Q-balls  (of $\omega=1.6$) sent with velocities $\,u=0.2$ and $\,u=0.05$ towards a hole of depth -0.9  located at $\,y_{0}=8.0$.}
\end{center}
\end{figure}

The magnitude and the type of the overall force between the Q-ball and the hole, when a Q-ball is sent with a certain speed and at a certain impact parameter, depends also on the charge of the Q-ball ({\it ie} $\omega$). Fig. 22 presents the trajectories of Q-balls
of different charges (different $\omega$)  when each Q-ball was sent with velocity 0.1 and towards a hole of -0.9 located at $\,y_{0}=8.5$. We note a very strong 
dependence on the charge. Sometimes the trajectories are deflected towards the hole, for 
other $\omega$ they are deflected in the other direction.
The amount of the deflection also depends strongly on $\omega$. Clearly this is a complicated process which requires further investigation.  One factor that certainly is partially responsible for this behaviour is the broadness of the profile functions. However,
this broadness is basically monotonic (but not at the tail (see fig. 14)) so it is hard 
to predict how much charge (in the form of `baby' Q-balls) get
generated in the hole and how this affects the overall scattering properties. All this behaviour requires more detailed further investigation.

 %The force increases as the charge increases and vice versa. Figure 8 shows the scattering trajectory of a Q-ball with $\omega=1.55$, which is very close to the thin wall approximation, was sent with a velocity of 0.1 to a hole located at $\,y_{0}=8.5$. The figure shows that the attractive force is great in this case.
\begin{figure}
\begin{center}
\includegraphics[angle=270, width=14cm]{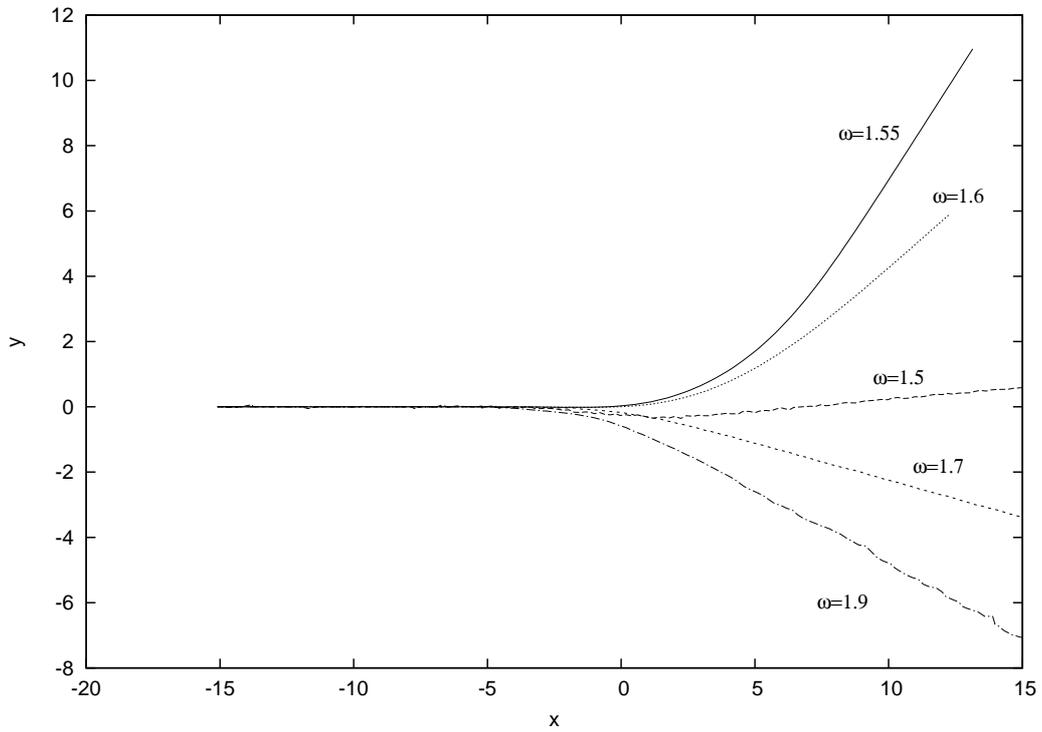}
\caption{The trajectories of a Q-ball with $\omega=1.5, 1.55, 1.6, 1.7, 1.9$  sent with $\,u=0.1$, a hole of -0.9 depth located at $\,y_{0}=8.5$ }
\end{center}
\end{figure}

Finally, we have also checked, numerically, that the forces are independent of the overall
helicity, that is the force does not depend on the sign of $\omega$. The behaviour is 
independent of whether we send a Q-ball or an anti-Q-ball at a hole  located at the same position. The dynamics also does not depend on the sign of the impact parameter; {\it ie}
it is also still the same when a Q-ball is sent (with the same velocity) towards a hole located at a corresponding negative impact parameter.

As the forces between a Q-ball and a hole
are attractive in the region of stability of the Q-ball all deflections can be cancelled
if we send a Q-ball symmetrically in-between two holes. Then the only effect is a small
attraction as the Q-ball approaches the line joining the holes followed by a repulsion.
 So, in this case the Q-ball behaves again like a point particle.

\section{\textbf{Conclusion}}

In this paper we have presented the results of our studies of the scattering of Q-balls on potential
obstructions (in the form of potential barriers and holes) in ($1+1)$ and $(2+1)$ dimensions.
In our work we have tried to compare our results with similar results for topological
solitons. And in many cases the results were similar; however, some results were different
as Q-balls have some conditions that they have to satisfy to be stable. Moreover, as their 
charge is not quantised they can, in the interaction with obstructions, generate
further small Q-balls (called `baby' Q-balls in this paper).

We have found that in the barrier case the scattering of Q-balls resembled very closely the scattering of topological solitons. The scattering was very elastic. For each barrier
there was a critical velocity below which the Q-ball was reflected and above which it was transmitted. What was surprising, at first sight, was that the value of this critical
velocity did not depend on the charge of the Q-ball ({\it ie} it was independent 
of $\omega$). As the scattering was very elastic we could estimate the value of this
critical velocity by performing an energy analysis ({\it ie} by comparing the energy
of a boosted Q-ball away from the barrier to the energy of Q-ball at rest at the top
of the barrier). The results were in an excellent agreement with what was seen in the numerical simulations. This suggests that, indeed, the scattering is very elastic and 
the approximation of Q-balls by point-like particles is quite reliable.

 For the scattering of Q-balls on potential holes the situation was slightly different as 
 the stable Q-balls, when they fall into a hole can become unstable.
 This instability translates into the generation of `baby' Q-balls which can then
 interact with their `parent'.
 
 We have performed detailed studies of the scattering of Q-balls on potential holes.
 When the Q-balls are stable, and remain stable in the hole, they, again, resemble 
 topological solitons. In this case they can either fall into the hole and get trapped 
 in it or pass through it with a slightly reduced velocity. The scattering is, again,
 quite close to being elastic although this time more radiation is produced than in the case
 of a barrier. However, when a Q-ball approaches the hole, especially when its 
parameters are close to the stability bounds, it produces small Q-balls in the hole.
As the basic force between Q-balls can be of either sign (it depends on their relative
phase) the `baby' Q-balls then start interacting with each other and with the
initial Q-ball.  The overall interaction can now be attractive or repulsive and this depends on the parameters of the configuration
(velocity and frequency of the Q-ball, the position and the height of the hole etc).

In ($2+1$) dimensions we have looked at the scattering of a Q-ball on a series of potential
  barriers whose position was varied along the $y$ axis. The Q-ball was deflected from the potential barrier. We have found that the deflection angle depended on the impact parameter, the height of the barrier, the velocity and the charge $\,Q$ of the Q-ball.
.

 We have also studied the interactions of stable Q-balls with the hole in ($2+1$) dimensions
at various impact parameters. We have found the expected deflection towards the hole although
the trajectories of the `parent' Q-ball were slightly irregular. This irregularity 
 increased at the lower impact parameters and it depended also on the velocity of the Q-ball. 
 When we studied the dependence of the impact parameter on the charge of the Q-ball
 ({\it ie} the frequency $\omega$) we have found a surprising irregularity as shown in fig. 22.
 For some frequencies the deflection was towards the hole, for some others away and it varied
 in magnitude in a very unpredictable way. Clearly the overall forces in this case are very finely balanced. It is clear that to get a good understanding
 of the scattering on a hole requires further work.

\textbf{Acknowledgement}
 One of us (JHA) would like to thank P.M. Sutcliffe for a helpful discussion.

\end{document}